\documentclass[preprint,aps]{revtex4}
\usepackage[latin9]{luainputenc}
\usepackage{float}
\usepackage{amsmath}
\usepackage{color}
\usepackage{amssymb}
\usepackage{graphicx}
\usepackage{setspace}

\makeatletter

\@ifundefined{textcolor}{}
{%
 \definecolor{BLACK}{gray}{0}
 \definecolor{WHITE}{gray}{1}
 \definecolor{RED}{rgb}{1,0,0}
 \definecolor{GREEN}{rgb}{0,1,0}
 \definecolor{BLUE}{rgb}{0,0,1}
 \definecolor{CYAN}{cmyk}{1,0,0,0}
 \definecolor{MAGENTA}{cmyk}{0,1,0,0}
 \definecolor{YELLOW}{cmyk}{0,0,1,0}
}

\usepackage{amsfonts}
\usepackage{bm}
\usepackage[dvips]{epsfig}
\usepackage[dvips]{graphicx}
\usepackage{float}
\usepackage{dcolumn}
\usepackage{ifpdf}
\usepackage{multirow}

\newcommand{\be}{\begin{equation}}
\newcommand{\ee}{\end{equation}}
\newcommand{\bes}{\begin{subequations}}
\newcommand{\ees}{\end{subequations}}
\newcommand{\ben}{\begin{eqnarray}}
\newcommand{\een}{\end{eqnarray}}

\makeatother

\begin{document}

\title{Bifurcation and chaos in one dimensional chains of small particles}

\author{Fabiano C. Simas$^{1,2}$, K. Z. Nobrega$^{3}$, and D. Bazeia$^4$}

\email {fc.simas@ufma.br;  kznobrega@ufc.br;  bazeia@fisica.ufpb.br}


\affiliation{
$^{1}$Programa de P\'os-Gradua\c c\~ao em F\'isica, Universidade Federal do Maranh\~ao
(UFMA), Campus Universit\'ario do Bacanga, 65085-580, S\~ao Lu\'is, Maranh\~ao, Brazil\\
$^{2}$Centro de Ci\^encias Agr\'arias e Ambientais (CCAA), Universidade Federal do Maranh\~ao (UFMA), 65000-000 Chapadinha, MA, Brazil\\
$^{3}$Departamento de Engenharia Teleinform\'atica, Universidade Federal do Cear\'a (UFC),
60455-640, Fortaleza, Cear\'a, Brazil\\
$^{4}$Departamento de F\'isica, Universidade Federal da Para\'iba (UFPB), 58051-970 Jo\~ao Pessoa, PB, Brazil}

\begin{abstract}
This work deals with bifurcation and the chaotic behavior in one dimensional chains of small particles. We consider two distinct possibilities, one where the particles are modeled by a fourth-order potential which was already studied. We revisit this system and bring novelties concerning the presence of the three-armed star behavior and the study of the corresponding Lyapunov exponent. The other system is new, and there the particles are more involved and modeled by an eighth-order potential. We investigate this new system within the same approach, emphasizing the behavior of the orbits, the bifurcation profile and the corresponding Lyapunov exponent.
\end{abstract}

\maketitle


\section{ Introduction }

The main interest of the present investigation is to study one dimensional chain of small particles, with the particles having internal structure and so making the system very attractive from the point of view of physical applications of practical interest. In the literature, there are several interesting studies related to the present investigation, and here we recall the following possibilities: if the small particles are magnetic, the related chain can be used to model magnetic quantum cellular automata, as explored in \cite{science} with the magnetic particles used to implement logic operations and propagate quantum information at room temperature. The one dimensional chain of small magnetic particles can also be used to describe the presence of fractal structures due to the nonlinear interactions among the particles, as studied in Ref. \cite{magnetic}. Moreover, the one dimensional chain can be used to investigate the metal-insulator transition observed in doped polyacetylene \cite{bak1} and also, the behavior of the nonlinear interactions leading to bifurcation and chaos, as described in Ref. \cite{bak2,bak3}.\\

\begin{figure}[h]
	\includegraphics[width=12cm]{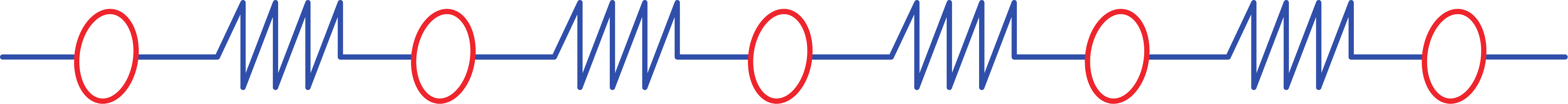} 
	\caption{The one dimensional chain of small particles.}
	\label{fig0}
\end{figure}

There are many other distinct motivations to study chaotic systems, some interesting possibilities have been discussed in \cite{bak3,book,Lya} and in references therein, including applications to Physics, Chemistry, Biology and Engineering. In this work, we shall treat the system of small particles as depicted in Fig. \ref{fig0}.  The blue springs are used to construct the one dimensional chain and they respond connecting the next neighbour in the usual manner. Its contribution to the Hamiltonian adds in the form of kinetic energy and has the standard form
\be 
T=\frac12 \sum_n (z_n-z_{n-1})^2.
\ee
In the above expression, $z_n$ identifies the $n$-th site of the chain and we have omitted a real parameter, in general used to weigh the kinetic contribution of the system. Here we use dimensionless quantities, so we will add the parameter $a$ to the potential energy, to do the task of weighing the potential contributions relative to the kinetic energy of the system. In the one dimensional chain, however, the red closed curves depicted in Fig. \ref{fig0} describe the small particles, and they then respond for the presence of nonlinear interactions in the model under consideration. The contribution to the one dimensional chain adds in the form of the potential energy. In order to specify the system, we have to identify its contribution to the chain, and an interesting standard nonlinear situation is in general described by a potential having the fourth-order power in the coordinate. It can be written as
\be\label{p4}
V=\frac14 a \sum_n (1-z_n^2)^2.
\ee
It is very important in physics and engenders the basics of the Ginzburg-Landau theory for a second-order phase transition \cite{Landau}, the Higgs mechanism that introduces spontaneous symmetry breaking that responds for the generation of mass for the elementary particles \cite{field}, the Kerr effect in nonlinear optics which allows for the presence of bright and dark solitons in fibers \cite{optics,APR} or the Landau-Ginzburg-Devonshire theory to describe negative capacitance in ferroelectric materials \cite{ferro,applied}. This potential is also present in Refs. \cite{magnetic,bak1,bak2,bak3}. It is also known as the double-well or $\phi^4$ potential, and has been used to describe other issues, in particular, movability properties of localized excitations \cite{PRL96} and thermal conduction \cite{thermal} in discrete one dimensional lattices, to quote two interesting possibilities.

In the present work we want to investigate the one dimensional chain depicted in Fig. \ref{fig0}, but we shall suppose the small particles have another internal structure, different from the above fourth-order potential. In fact, inspired by the work on kinks in scalar field theory with polynomial potential \cite{baz1}, here we shall study the eighth-order potential described by 
\be\label{p8}
V=\frac89\; a\; \sum_n \left(\frac14-z_n^2\right)^2(1-z_n^2)^2.
\ee
Differently from the fourth-order potential \eqref{p4} which engenders two distinct minima at each site, the potential \eqref{p8} has four minima, so it is more complex and describes a chain with small particles of different nature. Evidently, in this new chain the particles are somehow more complex then the previous ones and this will lead to distinct results, which we want to describe in this work. We implement the investigation organizing the study as follows: In Sec. \ref{sec2} we describe the main results, firstly reviewing the system described by the potential \eqref{p4}. However, we add new information, in particular, the presence of the three-armed star and the behavior of the Lyapunov exponent. We then go on and study the new system with the potential \eqref{p8}, carefully investigating orbits, bifurcations, starfish, three-armed star, the banana behavior and the Lyapunov exponent. We conclude the work in Sec. \ref{sec3}, adding some comments on the present investigation and describing new possibilities for future work.


\section{Results}\label{sec2}


The models to be considered in this work are described the by Hamiltonian
\be 
H=\frac12 \biggl[ \sum_n \left(\frac12(z_n-z_{n-1})^2+V(z_n) \right) \biggr].
\ee
As we have already informed, we are considering dimensionless quantities, so $z_n$ stands for the dimensionless position identifying the $n$-th small particle in the one dimensional chain, and $V(z_n)$ represents the corresponding potential. A sum in $n$ is considered to add all the particles. The equations of motion that follow from the above system can be obtained in the usual way. Here we follow the standard route considered in \cite{bak2} and we get the set of equations
\begin{eqnarray}
x_{n+1}  &=& - y_n + 2x_n + \frac{dV(x_n)}{dx_n},\nonumber \\
y_{n+1} &=& x_n,
	\label{var2}
\end{eqnarray}
where we have changed $z_n\to x_n$ and $z_{n-1}\to y_n$. Each model is now identified by specific potential, and below we consider the fourth-order and the eighth-order models separately.   


\subsection{The fourth-order model}

\begin{figure}
	\includegraphics[width=8cm]{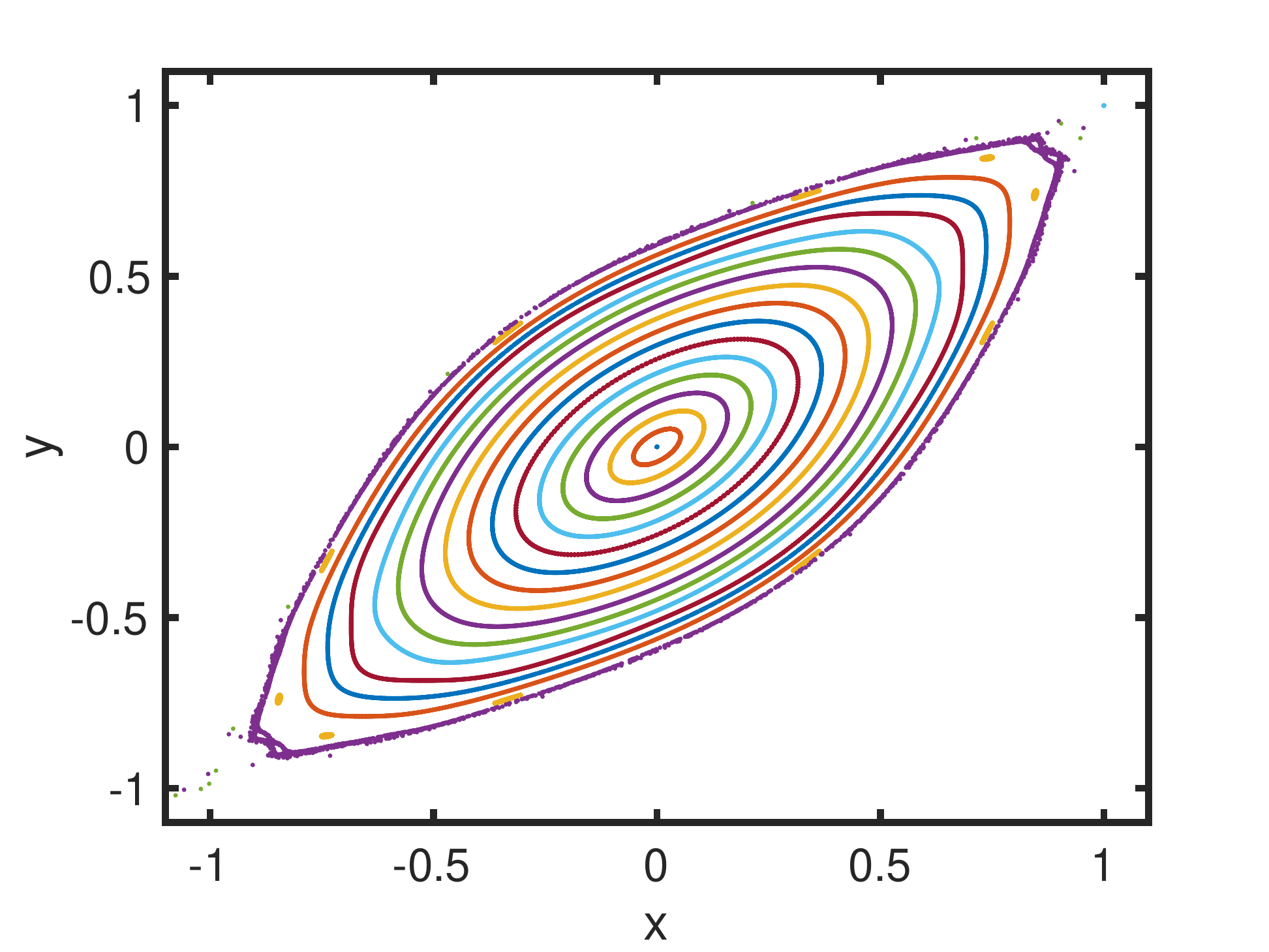} 
	\includegraphics[width=8cm]{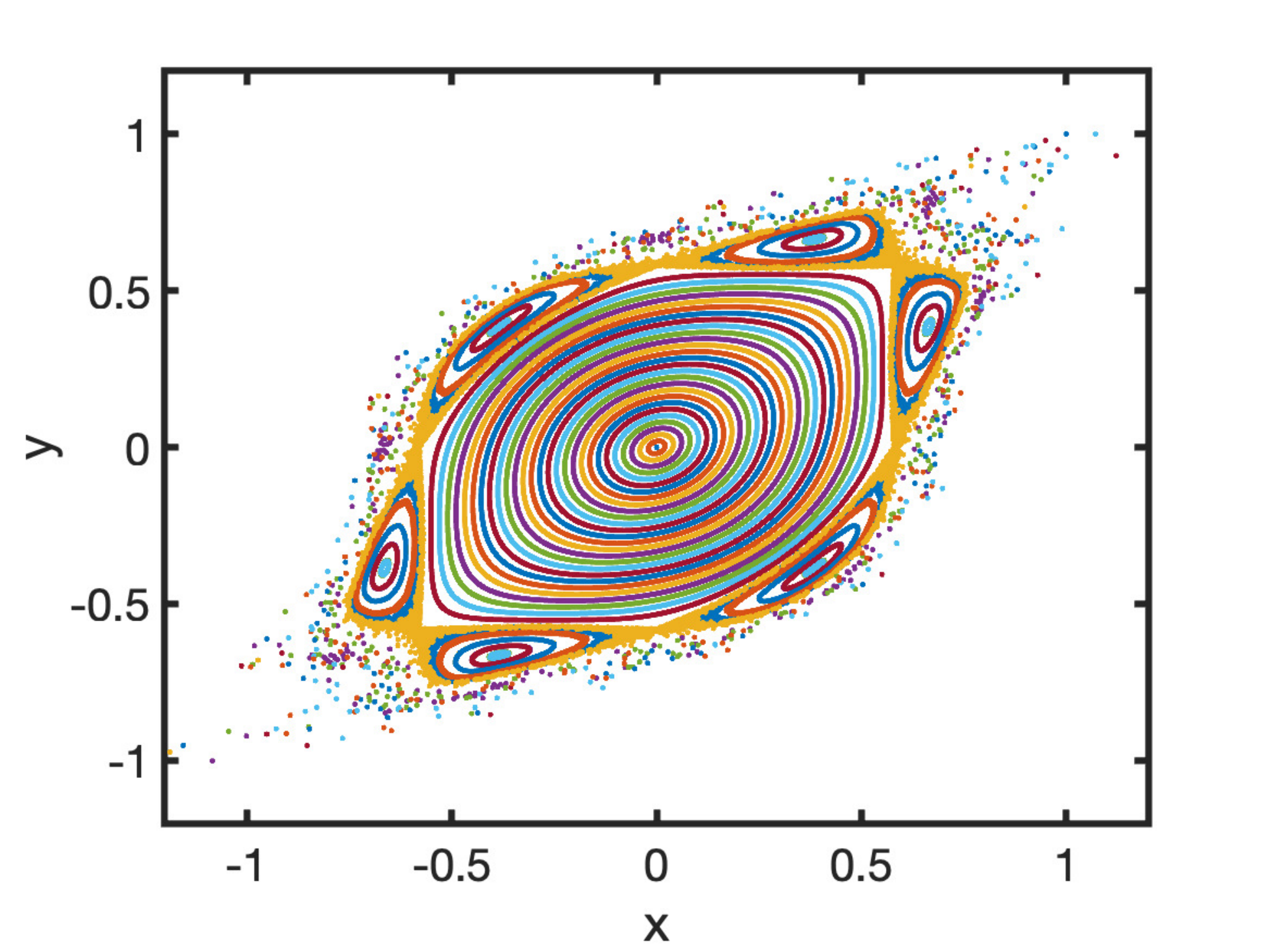} 
	\caption{(a) Orbits of the fourth-order model calculated for $a=0.8$ and (b) $a=1.5$.}
	\label{fig1}
\end{figure}

The fourth-order potential is specified by
\be   
V(x_n)=\frac14 a (1-x_n^2)^2,
\ee  
where $a$ is a real and dimensionless parameter that controls the potential energy of the system, in comparison with the kinetic energy. This potential is non-negative, has a local maximum at zero and two minima at ${\bar x}_n^\pm=\pm1$. It shows that the small particles in the chain have internal structure and this allows for the presence of nonlinear excitation in the chain, which we study below. The derivative of the potential to be used in Eq. \eqref{var2} has the form
\be\label{der4}
\frac{dV(x_n)}{dx_n}=-a\, x_n + a\, x_n^3.
\ee
Evidently, the driving parameter is $a$, and it measures the importance of the potential behavior compared to the kinetic contribution to the chain. In \cite{bak1}, infinite series of bifurcations were discovered. Using the mapping (\ref{var2}), the authors discovered an elliptical structure centered at the fixed point at the origin for $a={8}/{9}$. The same outcome can be found in the Ref. \cite{bak2}. We have also investigated the same system, and in Fig. \ref{fig1} we display the orbits for the fourth-order theory calculated for $a=0.8$ and $a=1.5$. This result demonstrates that for small values of $a$, there is just one fixed point at the origin.

The stability of fixed point can be studied by the tangent space orbits. We consider small fluctuations $x_n = \hat{x}_n + \delta x_n$ around $\hat{x}_n$; we implement this to get

\begin{eqnarray}
\left(
\begin{array}{c}
	\delta x_{n+1}\\
	\delta x_{n}  
\end{array} \right) = M_n \left(
\begin{array}{c}
	\delta x_{n}\\
	\delta x_{n-1} 
\end{array} \right),
\label{matrix}
\end{eqnarray}
where 

\begin{eqnarray}
M_n = \left(
\begin{array}{cc}
	3a\hat{x}_n-a+2 & -1\\
	1 & 0  
\end{array} \right).
\end{eqnarray}
The eigenvalues of $M_n$ are determined by the trace and determinant, in the form
\be
\frac{tr M_n}{2} \pm \sqrt{\frac{|tr M_n|^2}{4} - \det M_n}.
\ee
We realize that $\det M_n=1$, then the eigenvalues are solely determined by the trace \cite{bou1,gree1}. This allows us to determine whether the eigenvalues are real or not. The system will oscillate with non-real eigenvalues, consequently, the orbits will be stable under the following condition $|tr M_n| < 2$.

\begin{figure}
 	\includegraphics[width=8cm]{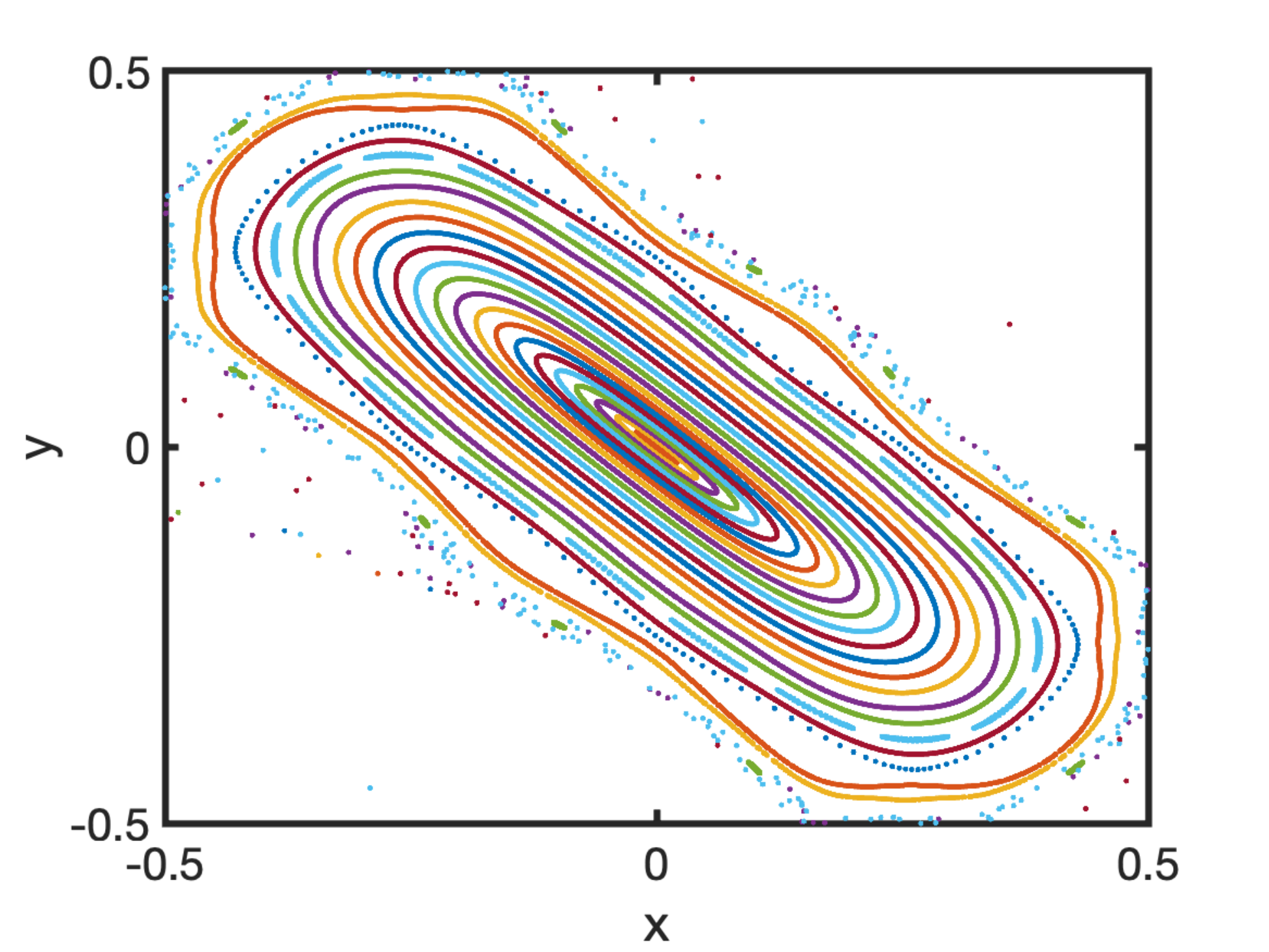} 
 	\includegraphics[width=8cm]{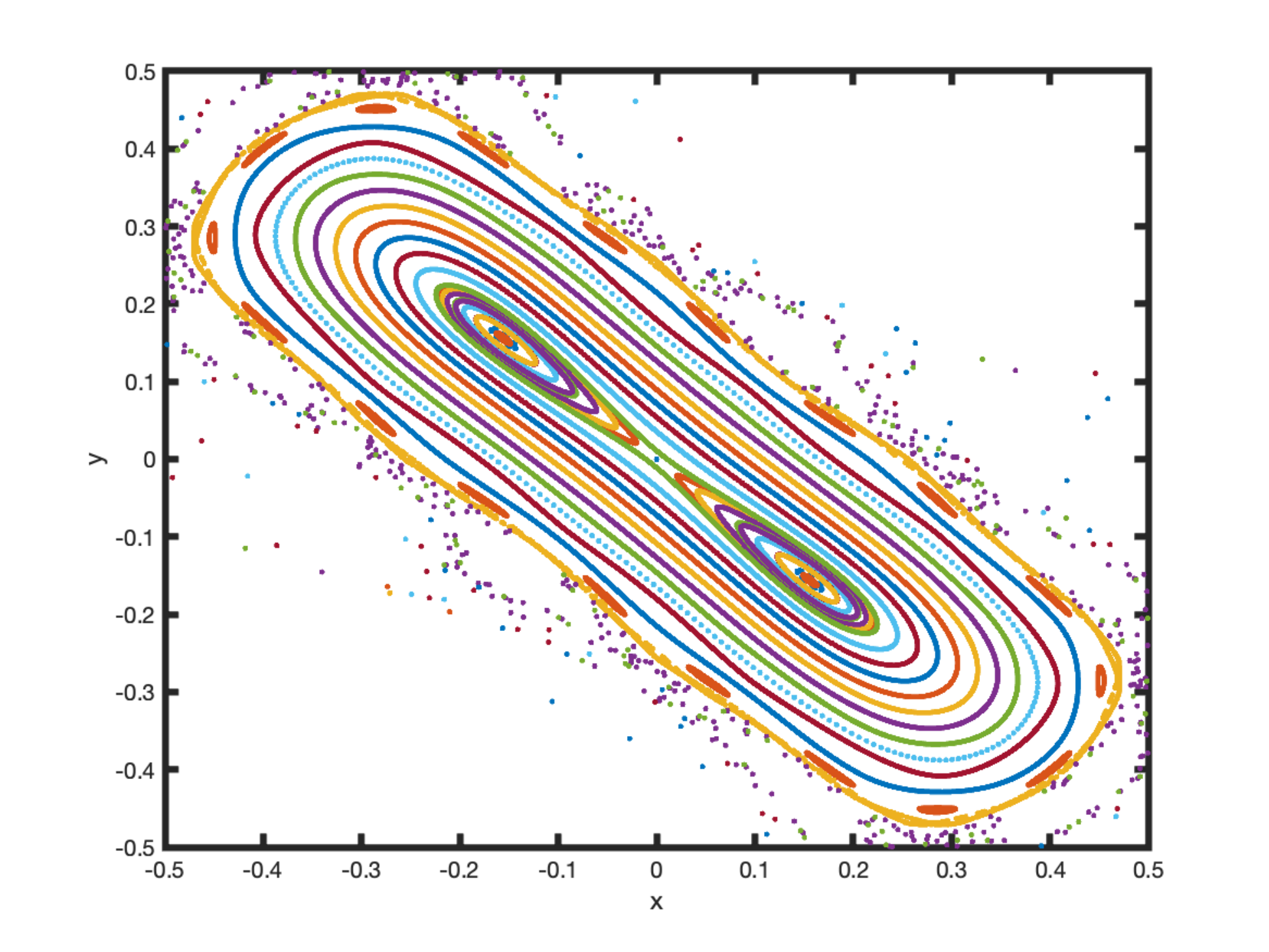} 
 	\caption{(a) Orbits of the fourth-order model calculated for $a=3.95$ and (b) $a=4.1$.}
 	\label{fig2}
 \end{figure}

As a result of investigation of stability, it has been shown that the fixed point at the origin becomes unstable for $a>4$. Therefore, in \cite{bak1} for values below the threshold $a=4$, bifurcations are impossible to observe. However, when one increases $a$ above $a=4$, the system bifurcates into $2$-cycle orbits. This is shown in Fig. \ref{fig2}, which displays the results for $a=3.95$ and $a=4.1$: in Fig. \ref{fig2}a, we observe the features of elliptic orbits around $(0,0)$; however, in Fig. \ref{fig2}b one shows that the original fixed point presents features of hyperbolic orbit. As a result, the fixed point around the origin gives rise to two new fixed points which are now encircled by orbits. This indicates that the original fixed point has become unstable, and a bifurcation has appeared, as it is illustrated in Fig. \ref{fig2}.

\begin{figure}
	\includegraphics[width=8cm]{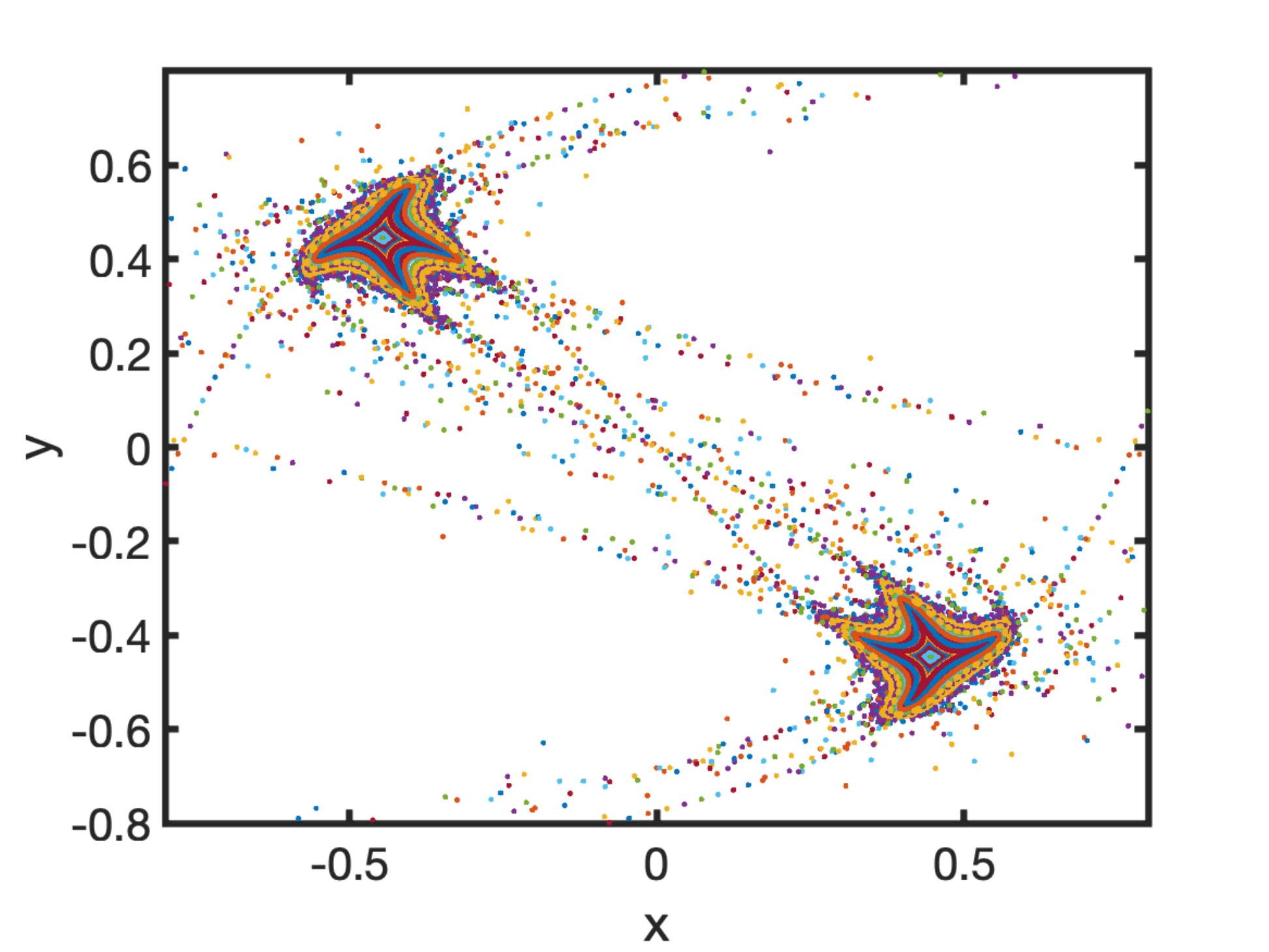} 
	\includegraphics[width=8cm]{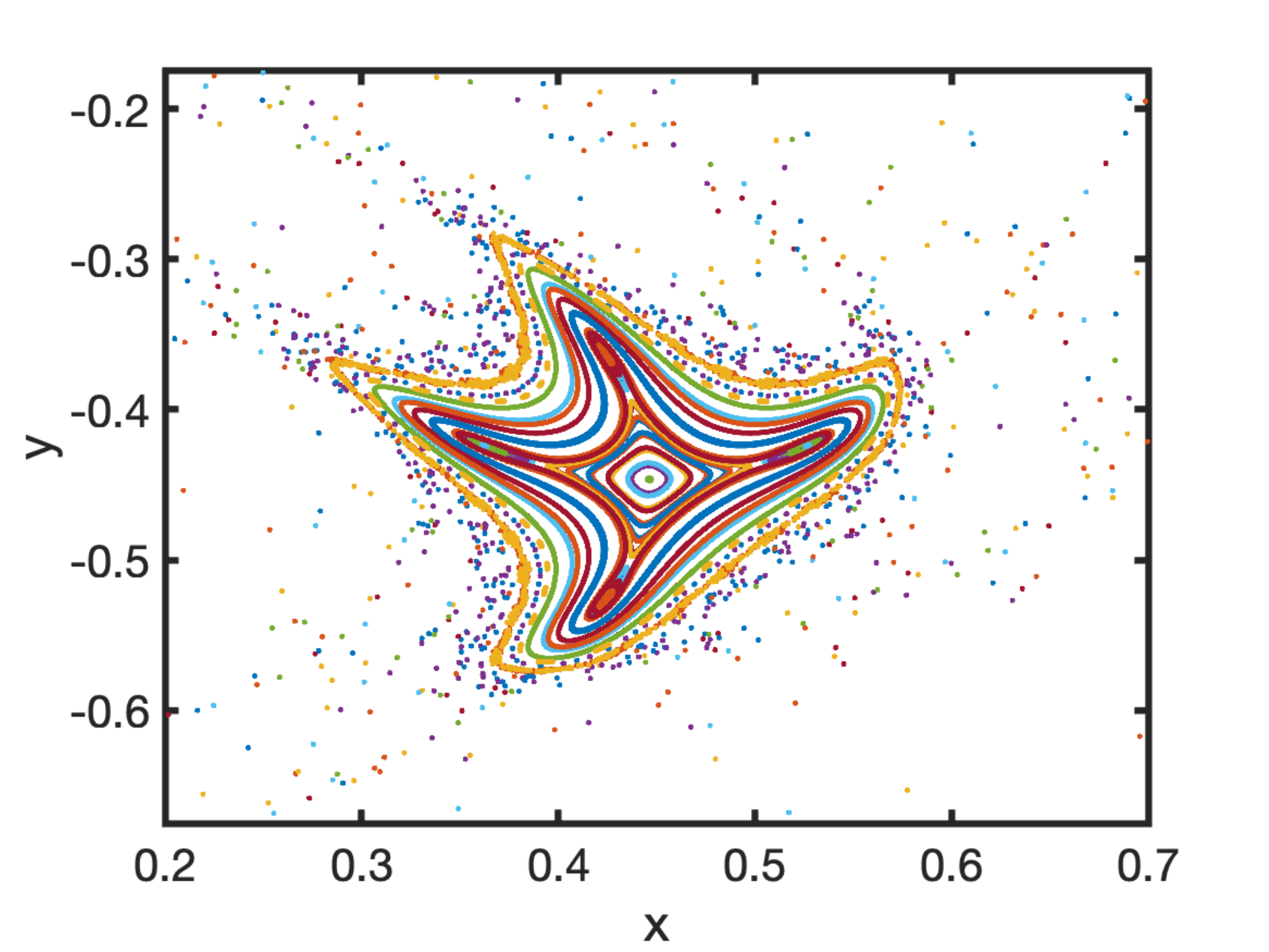} 
	\caption{(a) The starfish of the fourth-order model calculated for $a=4.995$ and (b) the zoom from figure (a) to $x>0$ and $y<0$.}
	\label{star}
\end{figure}

 With the increase of the value of parameter $a$, another fascinating result emerges. The 2-cycle orbits become slightly unstable at $a=5$. However, no bifurcations are envisaged \cite{bak2}. The Fig. \ref{star} show the result of computer iterations for $a=4.995$. In the Fig. \ref{star}a, the entire pattern of behavior is visible. In Fig. \ref{star}b, we observe a zoom in on the positive component of $x$. The Fig. \ref{star} depicts the starfish behavior identified before in Ref. \cite{bak2}. The regular orbit eventually deforms as $a$ becomes larger ($a>5$), resulting in an irregular starfish.

\begin{figure}
	\includegraphics[width=8cm]{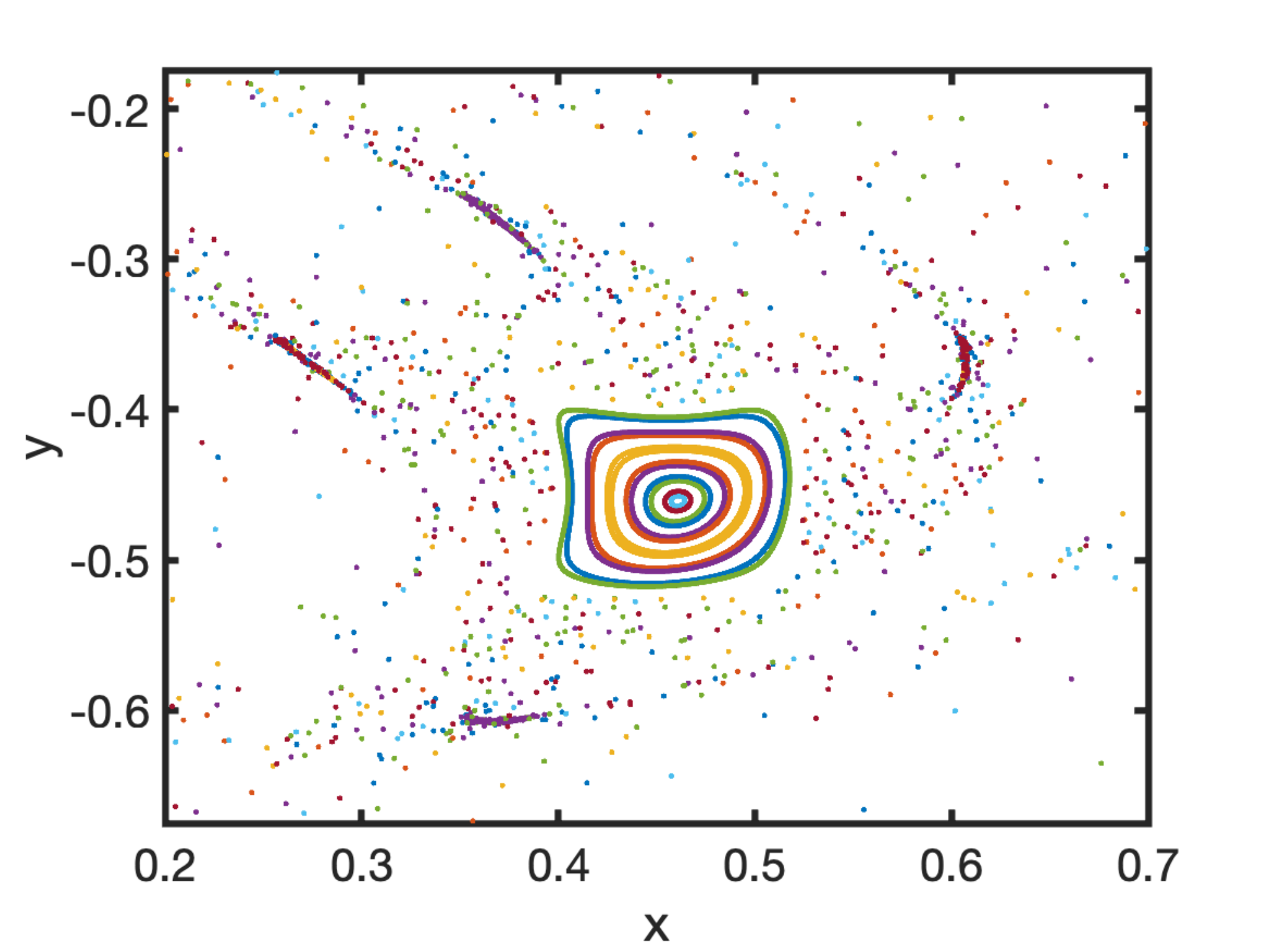} 
	\caption{Orbits of the fourth-order model calculated for $a=5.078$.}
	\label{defor}
\end{figure}

\begin{figure}
	\includegraphics[width=8cm]{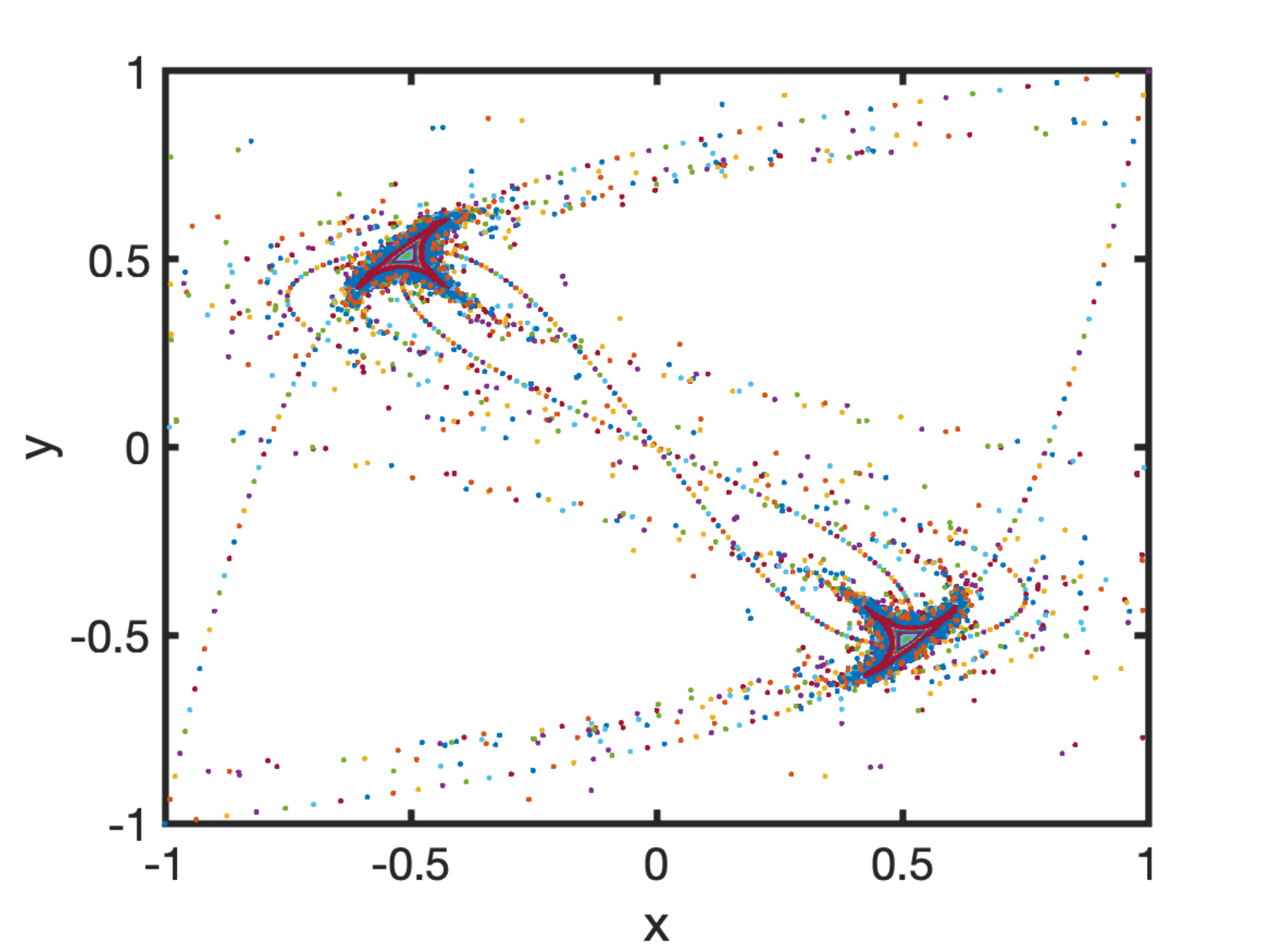} 
	\includegraphics[width=8cm]{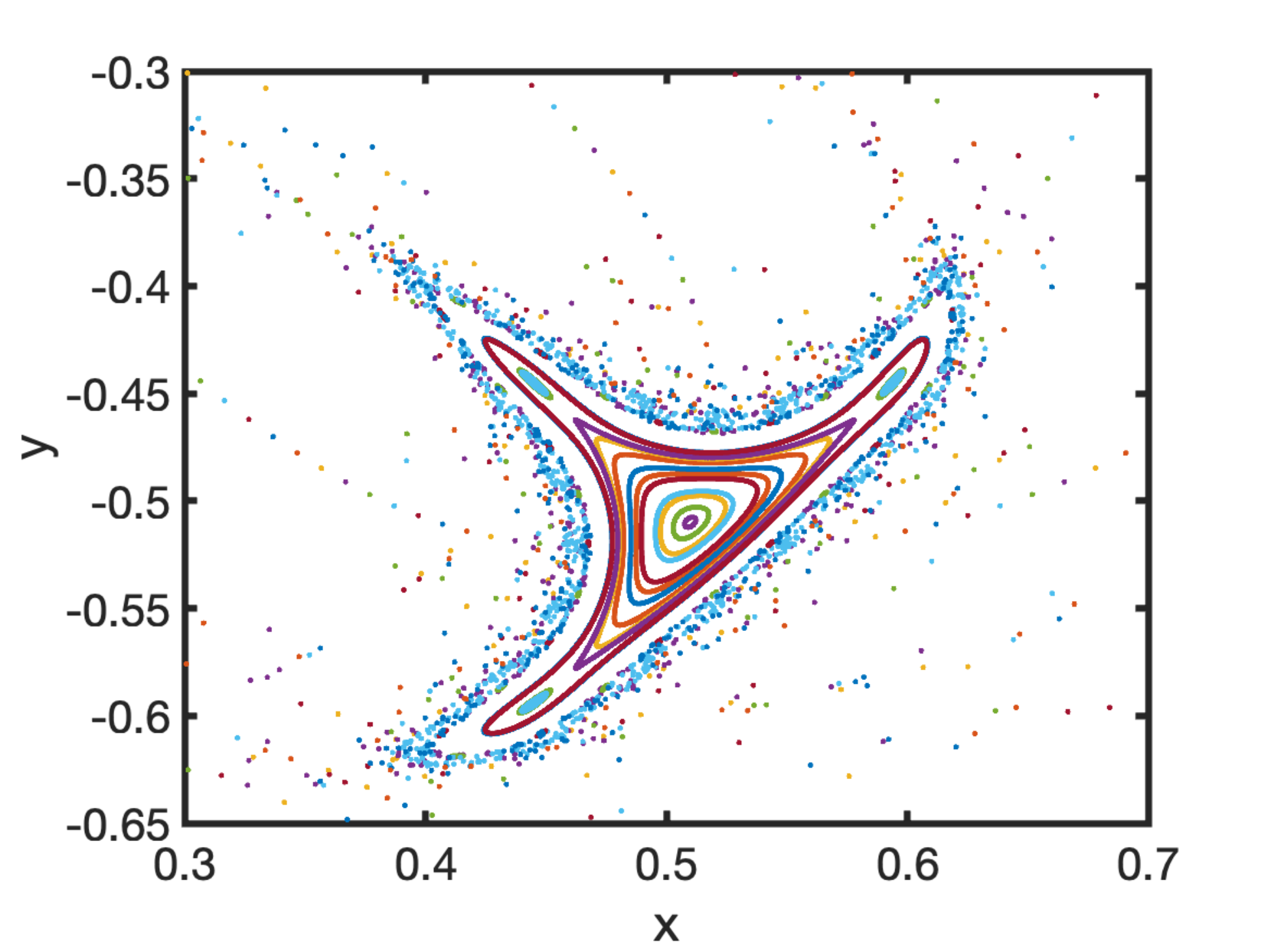} 
	\caption{(a) Orbits of the fourth-order model calculated for $a=5.405$ and (b) the zoom from figure (a) to $x>0$ and $y<0$.}
	\label{satr2}
\end{figure}

The starfish is no longer visible for $a=5.078$ because its arms disappear, as seen in Fig. \ref{defor}. The orbits that identified the starfish slowly disappear as we increase the parameter $a$. As illustrated in Fig. \ref{satr2}, however, the increasing of $a$ resulted in the identification of a novel behavior for the fourth-order model. For $a=5.405$, we see the creation of three arms, each with its own island. Fig. \ref{satr2}a depicts the whole pattern, whereas Fig. \ref{satr2}b displays a zoom to the positive component of $x$. This was not noticed in \cite{bak2}, the presence of the three-armed star depicted in Fig. \ref{satr2}, in contrast to the starfish in Fig. \ref{star}. It is important to observe that there is no bifurcation since the fixed points, like in the starfish, are surrounded by orbits.

\begin{figure}
	\includegraphics[width=8cm]{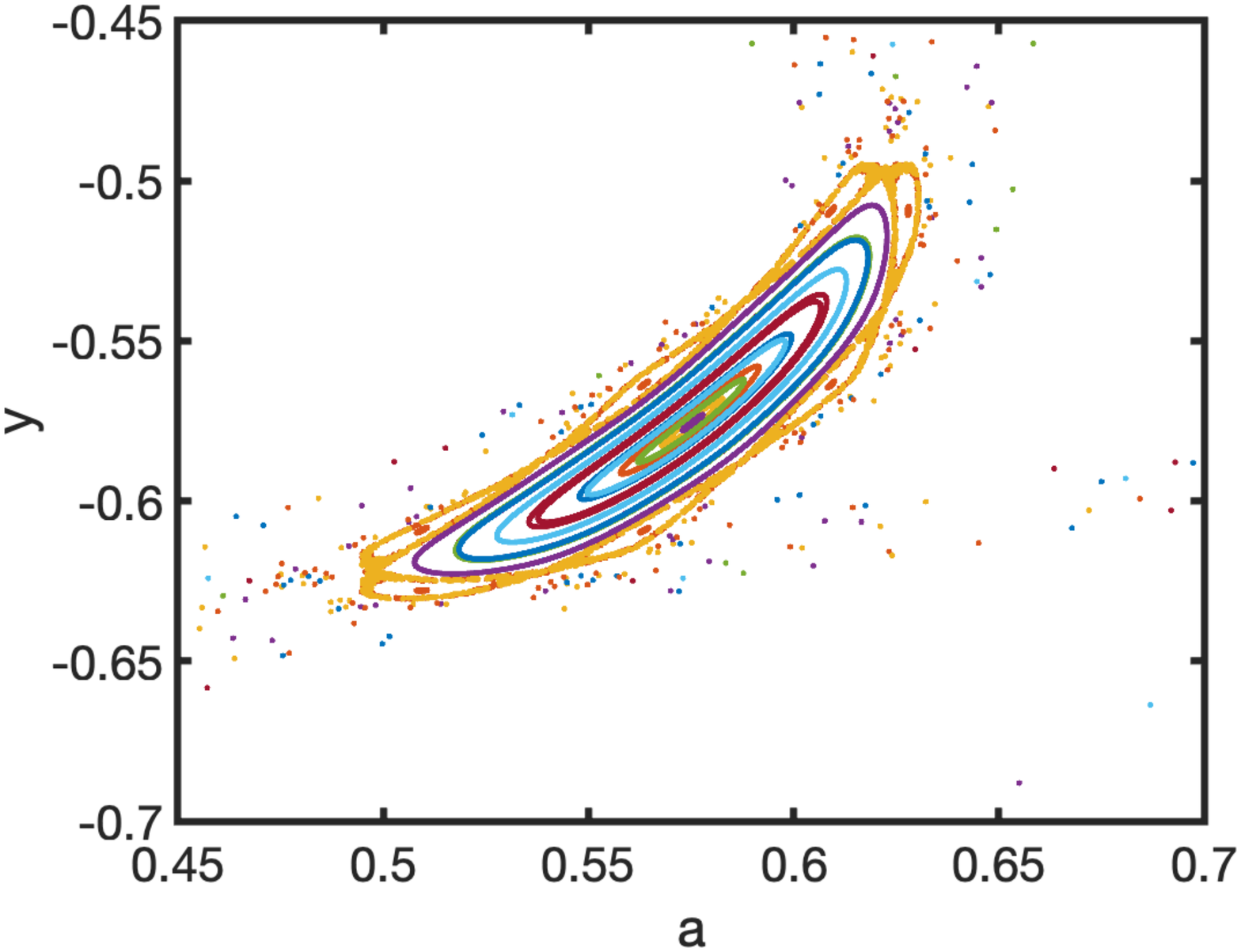}
	\includegraphics[width=8cm]{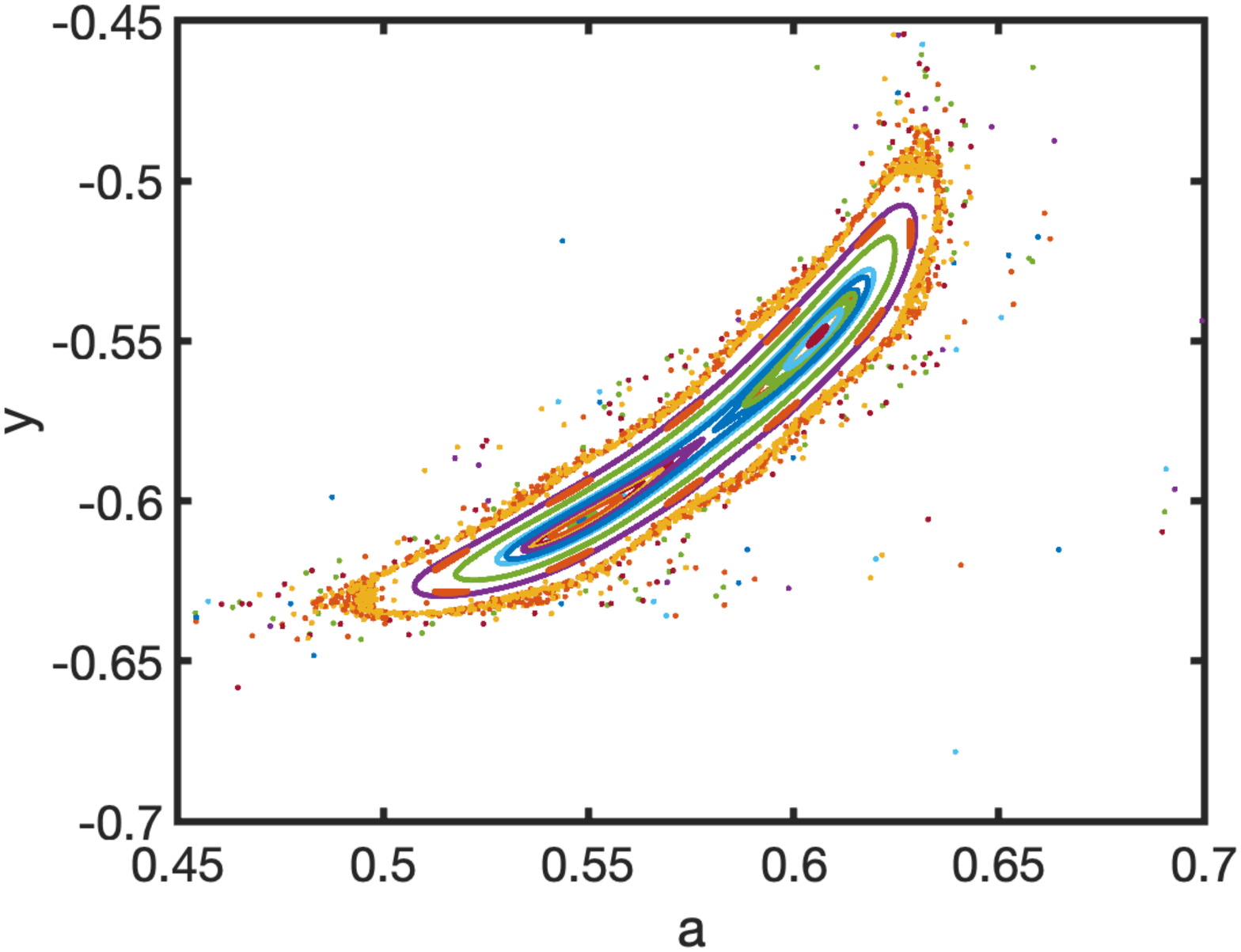} 
	\includegraphics[width=8cm]{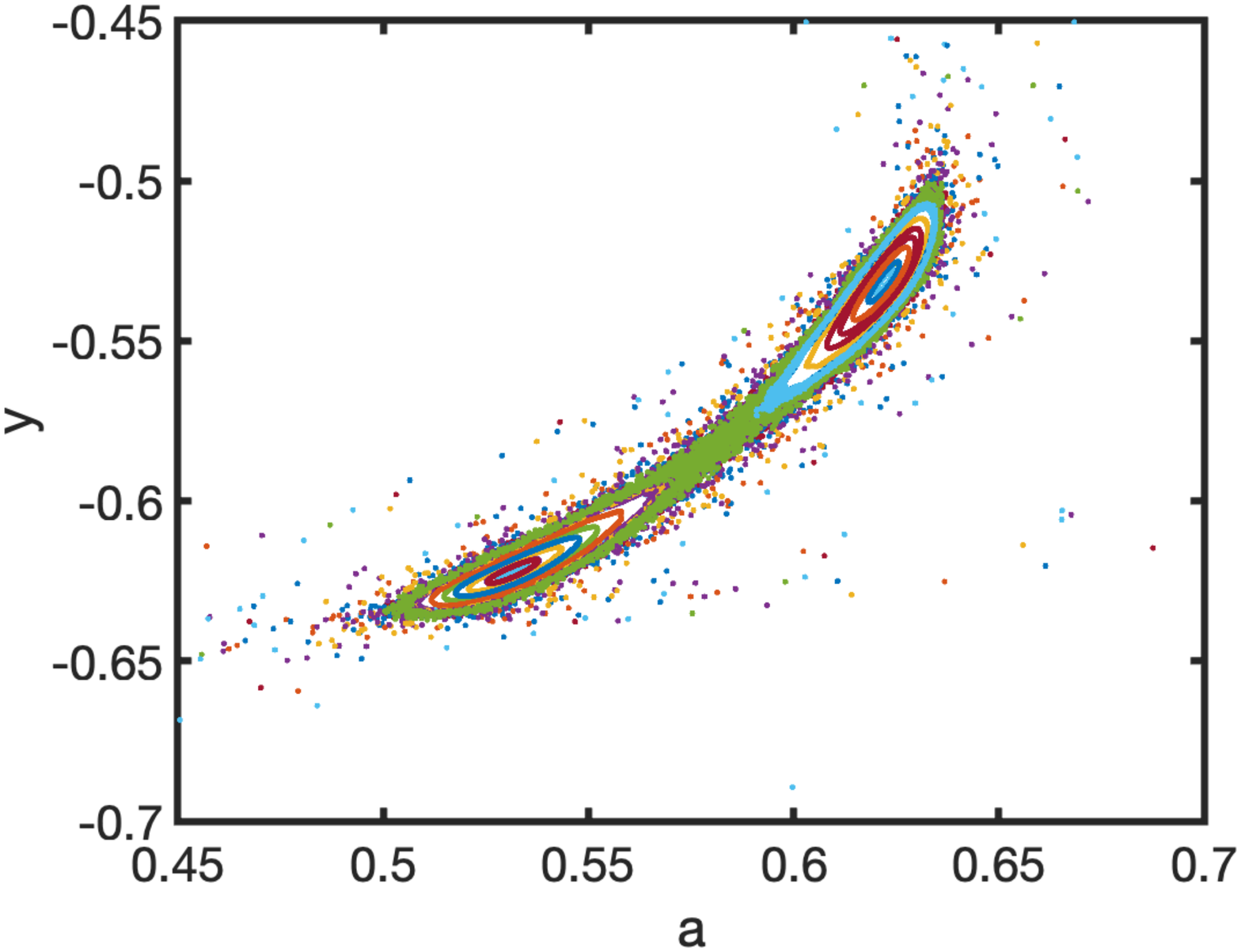} 
	\caption{The banana behavior of the fourth-order model calculated for (a) $a=5.98$, (b) $a=6.02$ and (c) $a=6.05$.}
	\label{ban}
\end{figure}

We notice that as the parameter $a$ is increased, the starfish disappears, becoming the three-armed star which also changes to become the banana structure found before in Ref. \cite{bak2}. The orbits calculated for $a=5.98$ is shown in Fig. \ref{ban}a. There is no bifurcation at this point, and the fixed point is surrounded by curves. However, when $a>6$, the initial fixed point becomes hyperbolic and, consequently, unstable. The banana becomes a banana split at $a=6.02$ in Fig. \ref{ban}{b}, and it further splits in the form depicted in Fig. \ref{ban}c for $a=6.05$.

The final feature of the fourth-order model that we consider is the connection that exists between bifurcation and chaos. We observe an elliptic orbit around the fixed point for smaller values of $a$. As this value increases, two new orbits are formed and the original fixed point becomes unstable. Then, the process of forming new orbits repeats itself as this parameter increases, until the system becomes chaotic.

\begin{figure}
	\includegraphics[width=8.0cm]{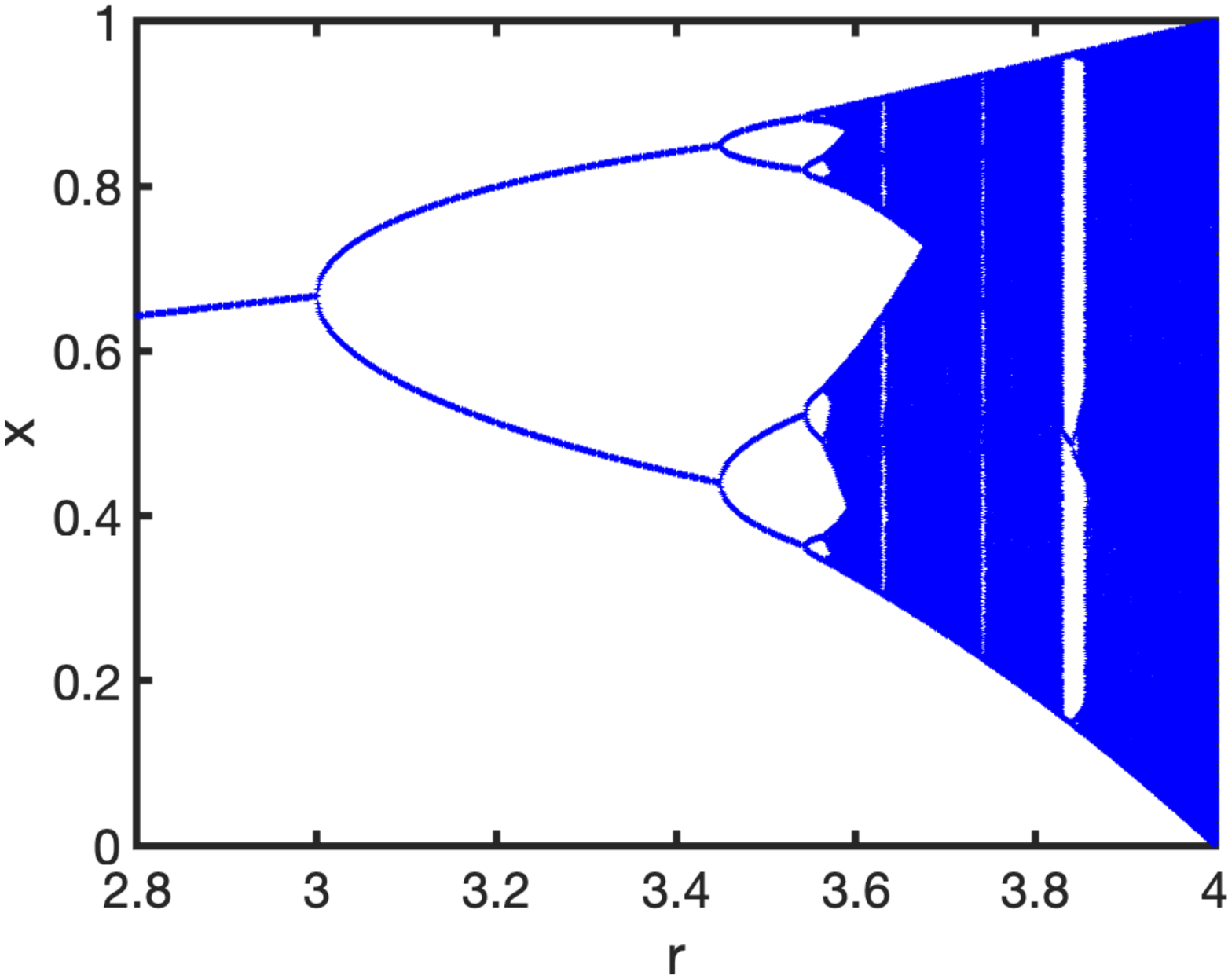} 
	\includegraphics[width=8.0cm]{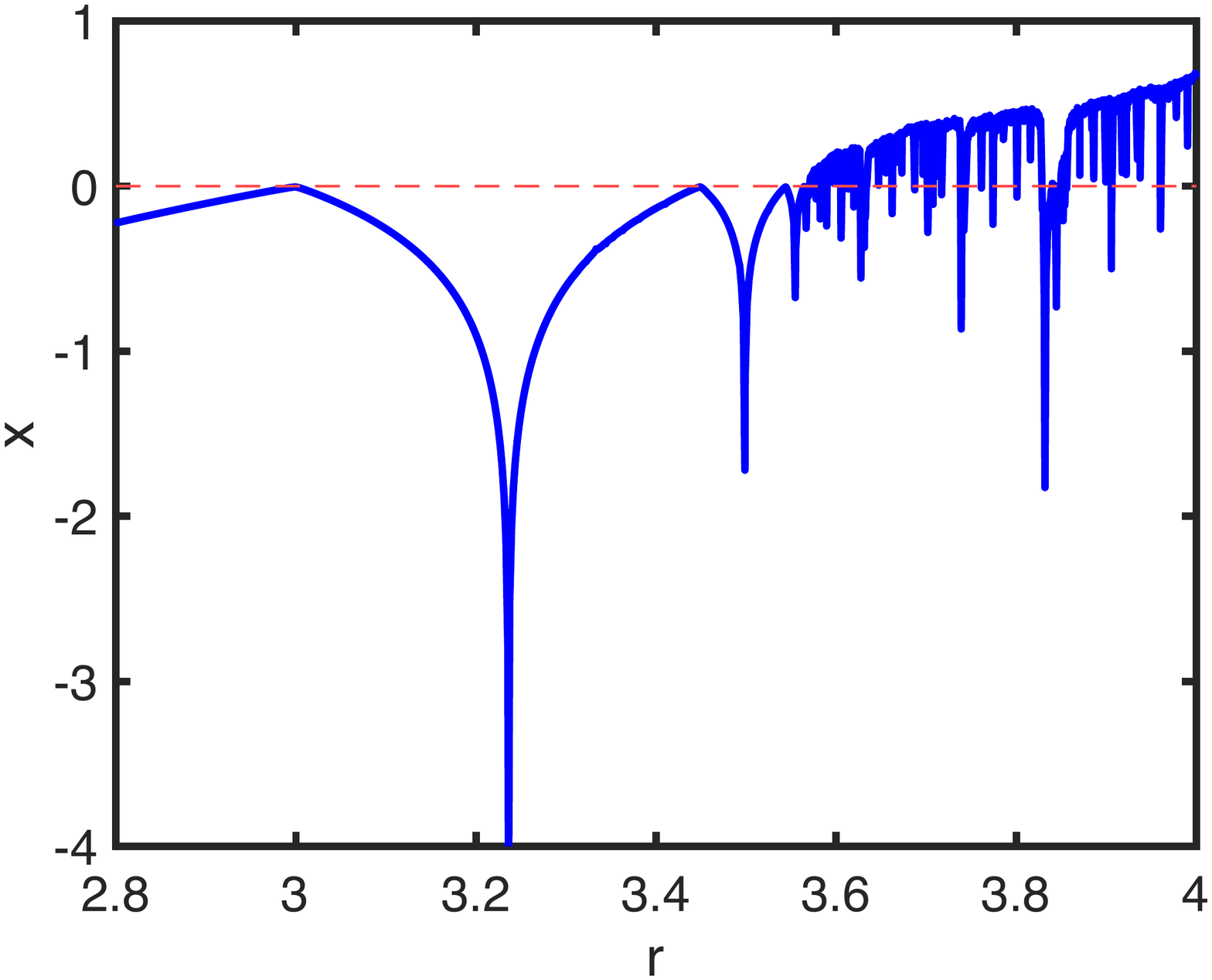} 
	\caption{(a) Bifurcation diagram of the logistic map and (b) corresponding behavior of Lyapunov exponent.}
	\label{logmap}
\end{figure}

\begin{figure}
	\includegraphics[width=8.0cm]{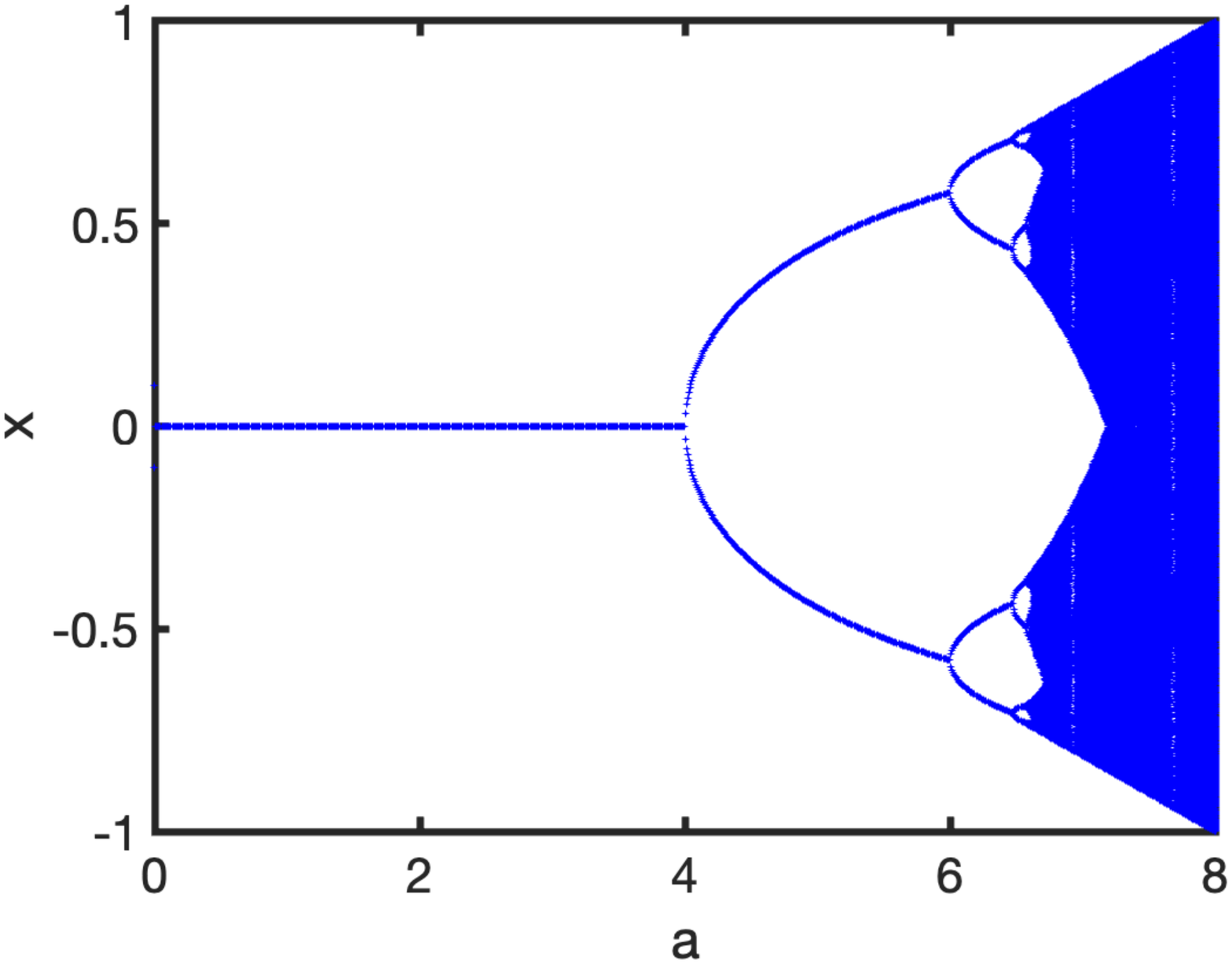} 
	\includegraphics[width=8.0cm]{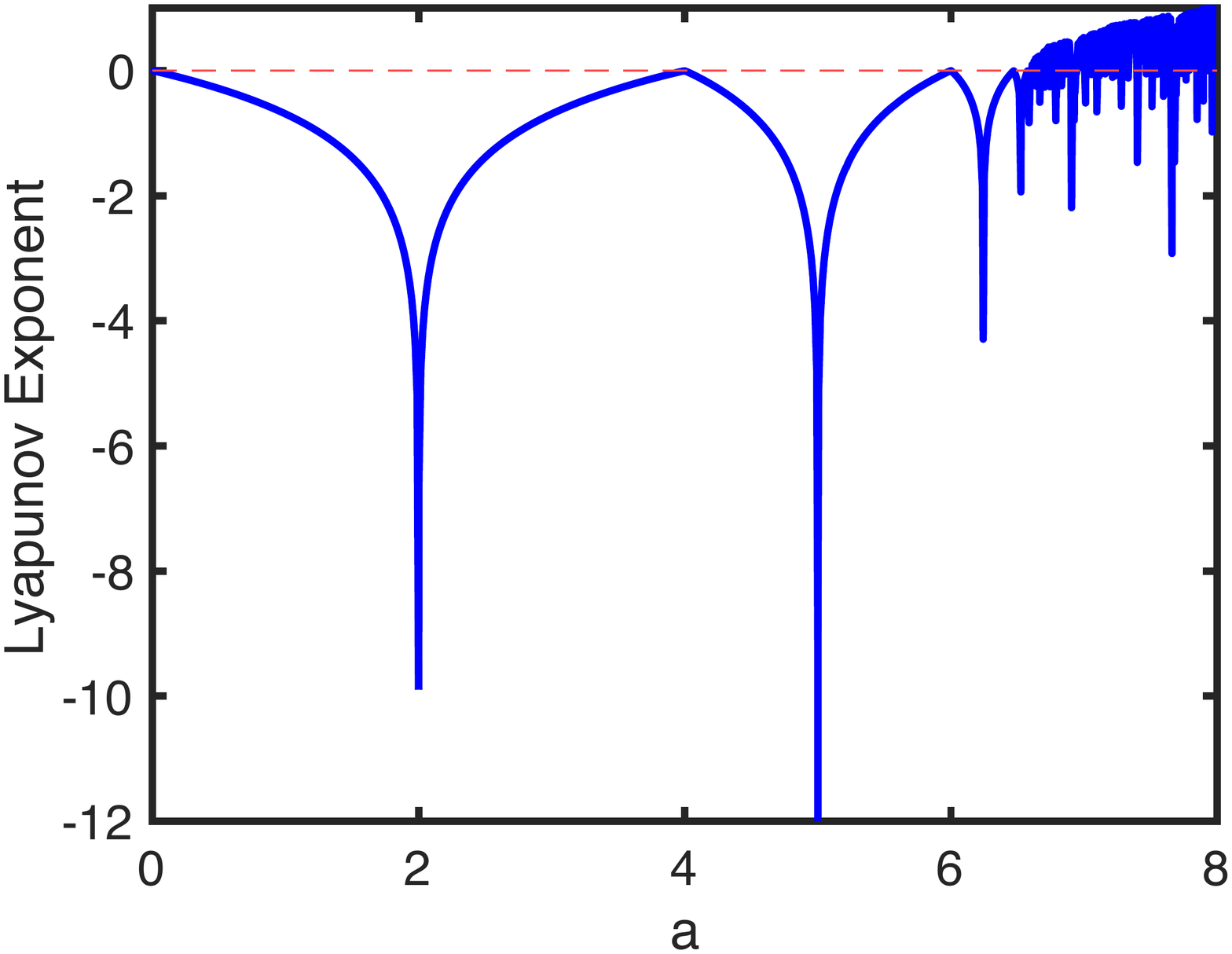} 
	\caption{(a) Bifurcation diagram of the fourth-order model and (b) corresponding behavior of Lyapunov exponent.}
	\label{lya}
\end{figure}

To inspect the behavior of the Lyapunov exponent we use Refs. \cite{book,Lya}. We first illustrate the subject investigating the famous logistic map, which is described by $x_{i+1}=r x_i(1-x_i)$, where $r$ is a real parameter know as the growth rate. In this case, we can calculate the bifurcation diagram and Lyapunov exponent standardly. The results are displayed in Fig. \ref{logmap}, and reproduces the corresponding results described in Ref. \cite{book}. These are well-known results, and we notice that for $r<r_{\infty} = 3.569$ the Lyapunov exponent is non-positive. Also, for $r$ small, in the region before reaching the chaotic behavior, one identifies the values $r=3$, $r=3.449$ and $r=3.544$ where the Lyapunov exponent approaches zero; these values are also seen in the bifurcation diagram, and there they identify the first, second and third bifurcation, respectively. This behavior also appears in the one dimensional chain with fourth-order potential. To see how it appear, let us now turn attention to the presence of bifurcation and chaos in the fourth-order model. We investigate both the bifurcation diagram and the Lyapunov exponent and display the results in Fig. \ref{lya}. There one sees that the Lyapunov exponent is non-positive for $a$ smaller than $a_{\infty} = 6.605$. Also, it approaches zero at $a=4$, $a=6$ and $a=6.472$, before the system engenders chaotic behavior. As expected, these values also identify the presence of the first, second and third bifurcation in the bifurcation diagram in Fig. \ref{lya}a.

In addition, there are minima in the figure, as seen at $a=2$ and $a=5$. The minimum at $a=2$ corresponds to the boundary region, where from it, the elliptical orbit rotates $90^{\circ}$. In order to view this, contrast Fig. \ref{fig1}a for $a=0.8$ ($a<2$) with Fig. \ref{fig2}a for $a=3.95$ ($a>2$). Finally, for large values of $a$, the Lyapunov exponent becomes positive, indicating the appearance of chaos.

Although some of the above investigation are revisions of previous results, as described in Ref. \cite{bak2}, for instance, the appearance of the three-armed star and the study of the Lyapunov exponent are novel, and confirm the chaotic behavior of the system for larger and larger values of $a$. We have found another work, Ref. \cite{cama}, where the authors implement similar investigations, using another model, with the same double-well profile, but with the interactions being controlled by hyperbolic functions. However, the investigation does not study the Lyapunov exponent.

\begin{figure}
	\includegraphics[width=8.0cm]{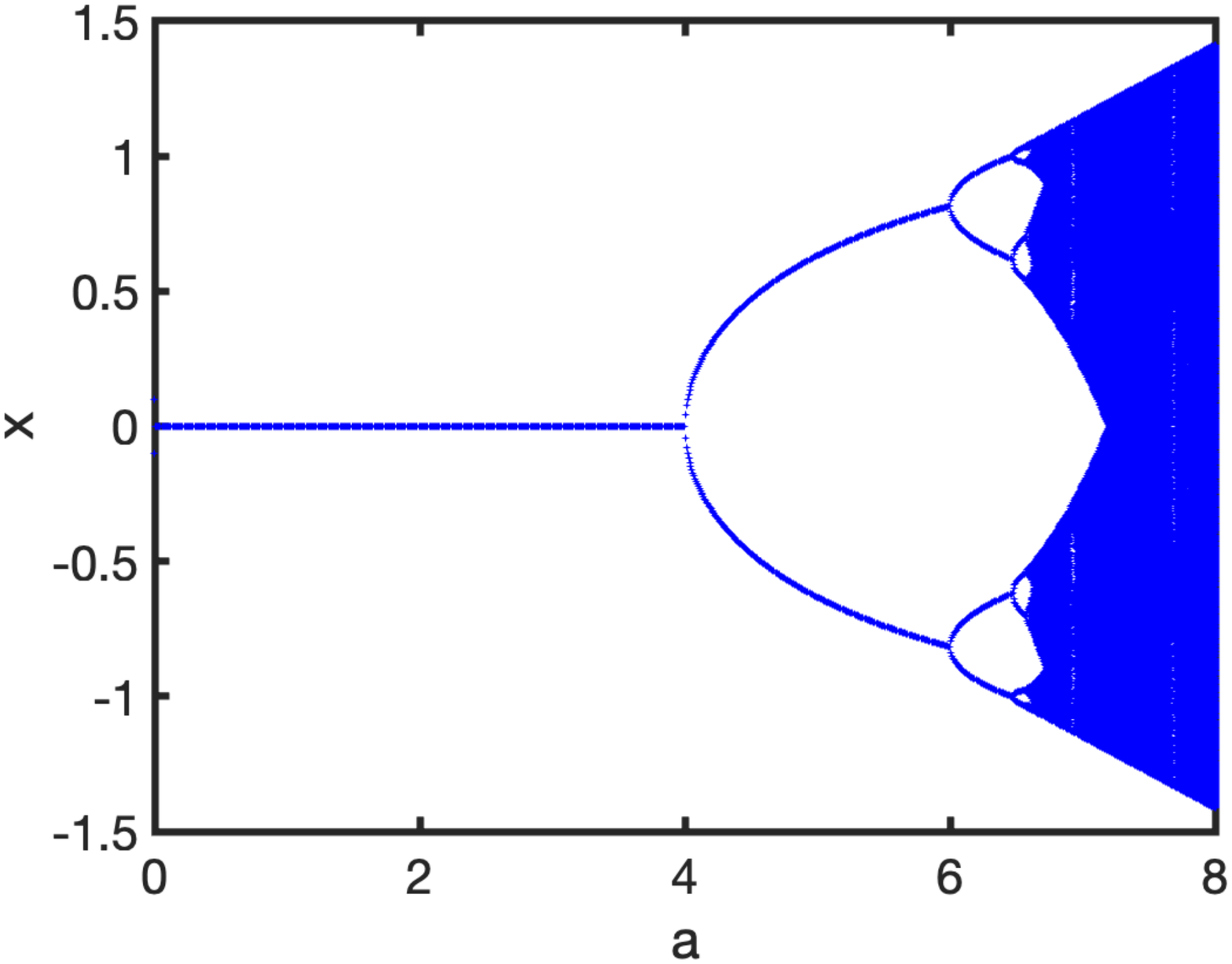} 
	\includegraphics[width=8.0cm]{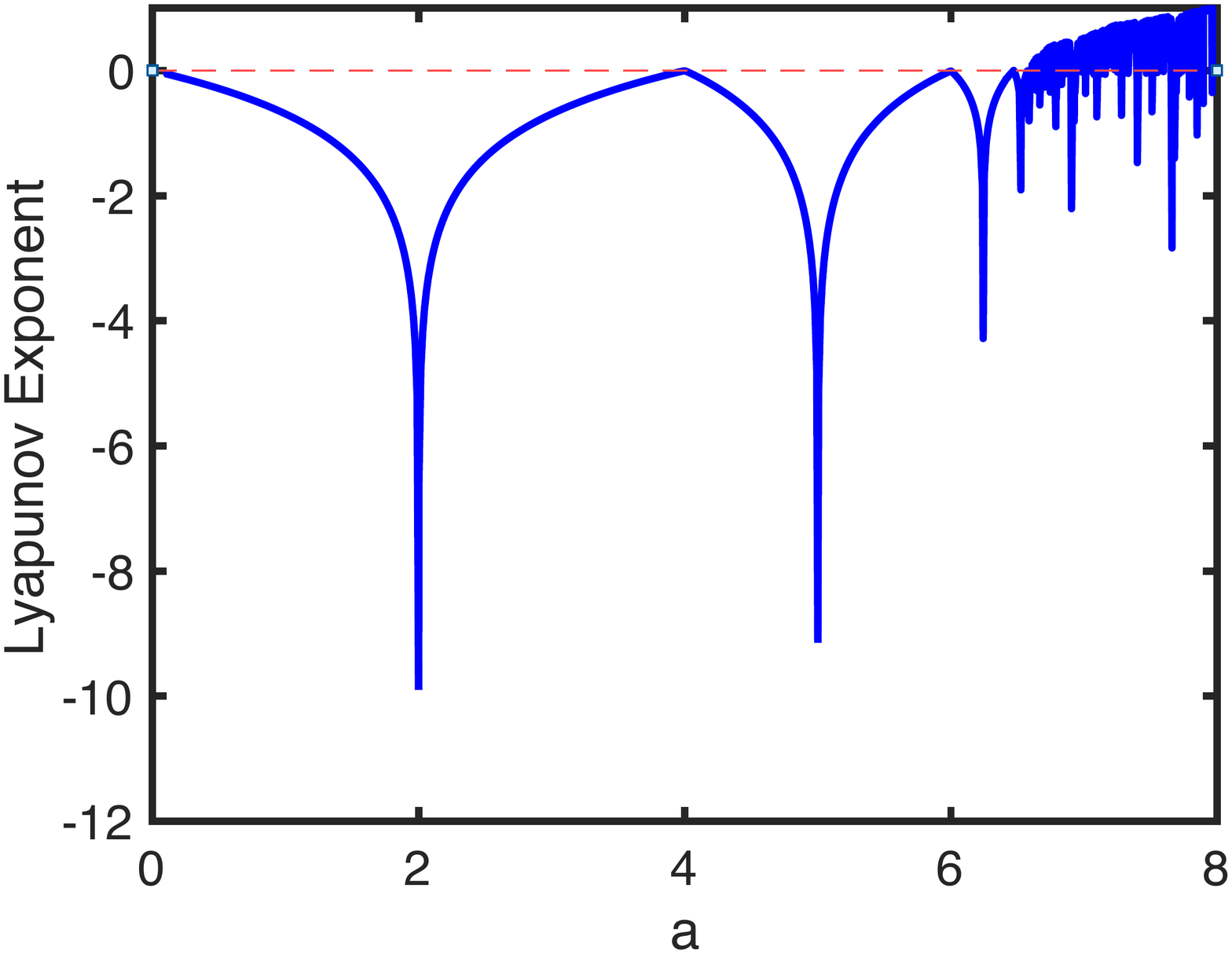} 
	\caption{(a) Bifurcation diagram of the fourth-order model and (b) corresponding behavior of Lyapunov exponent, for $b=0.5$.}
	\label{figX}
\end{figure}

In order to better understand the importance of the non-linear interaction in the model and also, to check the robustness of the numerical investigation that we developed above, we have considered a similar study, with the inclusion of another real and positive parameter, $b$, to control the weight of the interaction; that is, we have changed the fourth-order potential in a way such that the derivative in Eq. \eqref{der4} changes to
\be
\frac{dV(x_n)}{dx_n}=-a\, x_n + a\,b\, x_n^3,
\ee
where $b$ can be considered small or large, compared to $b=1$ which we have just considered. We have developed several calculations, for $b$ smaller and larger than unity, and we have identified the same qualitative behavior. To illustrate this, in Fig. \ref{figX} we display both the bifurcation and the Lyapunov exponent, for $b=0.5$, for instance. We have also checked that as we diminish $b$, the fixed points separate from each other; this is visible when we compare the bifurcation diagrams in Figs. \ref{lya}a and \ref{figX}a. Moreover, in the limit $b\to0$ the system trivializes, as expected. In Fig. \ref{figX}, we also notice the correct correspondence between the first, second and third bifurcations and the three first zeroes of the Lyapunov exponent, as expected.

\subsection{The eighth-order model}

\begin{figure}
	\includegraphics[width=8cm]{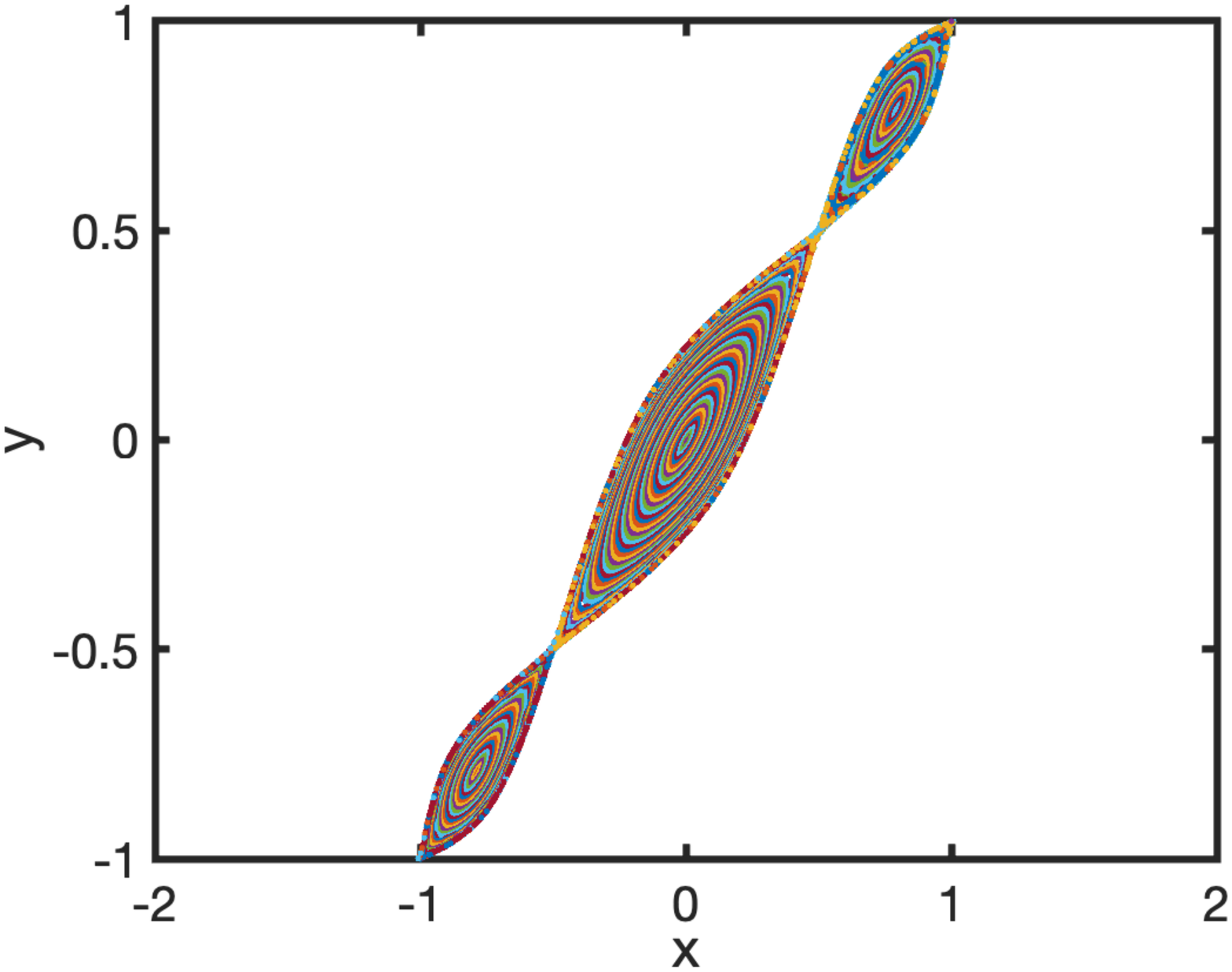} 
	\includegraphics[width=8cm]{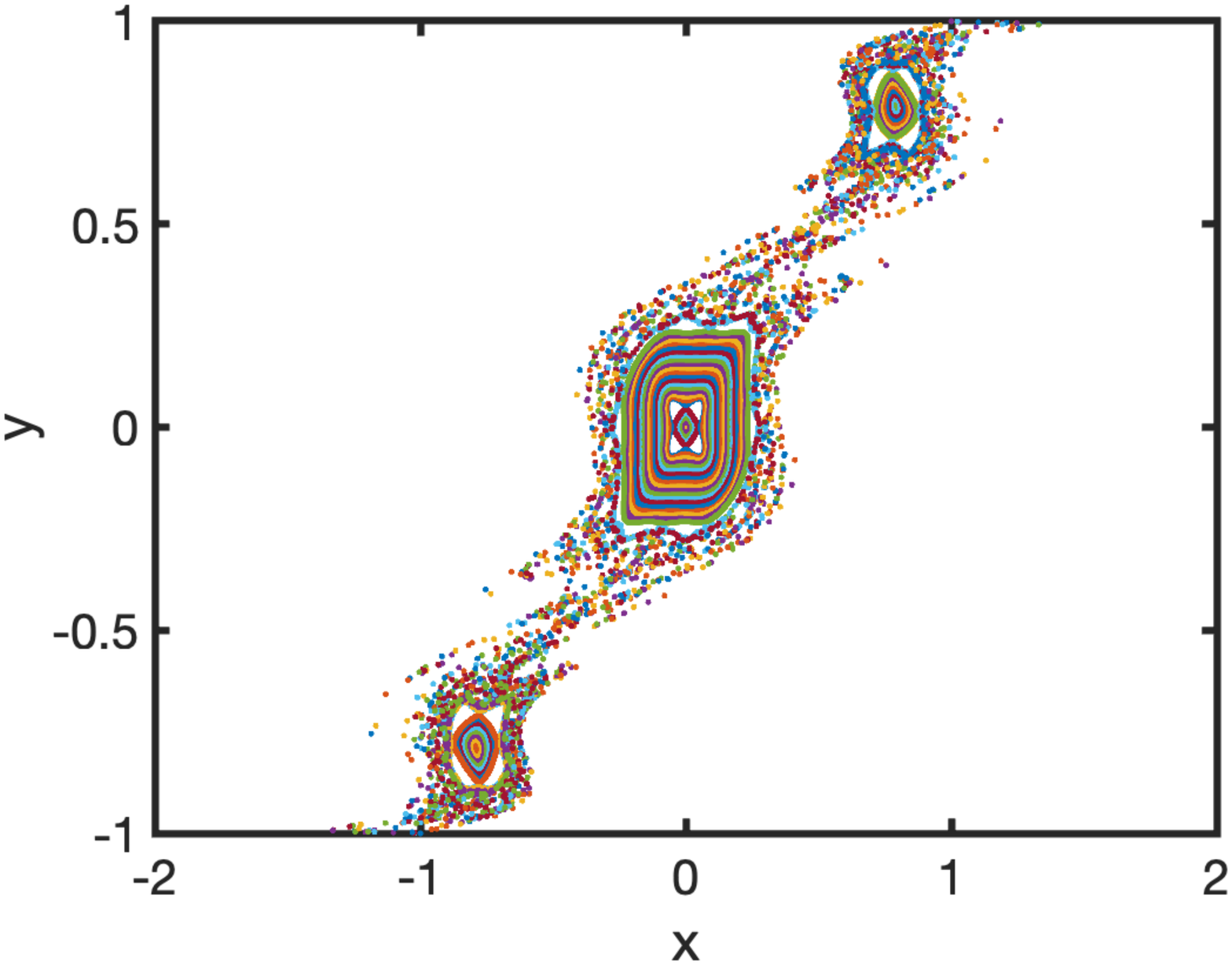} 
	\caption{(a) Orbits of the eighth-order model calculated for $a=0.5$ and (b) $a=1.83$.}
	\label{fig3}
\end{figure}

Let us now investigate the eighth-order potential shown in Eq. \eqref{p8}. It is inspired by Ref. \cite{baz1}, where it was shown that in the continuum, the relativistic real scalar field model is capable of supporting kinklike solutions in the distinct topological sectors present in the model. Since it has a richer structure, when compared to the previous fourth-order model, it is then worth examining its discrete version under similar procedure. The potential in Eq. (\ref{p8}) has three local maxima, one at the origin and two other, symmetric, at $\pm0.79$. Also, it has four global minima, two at $\pm 0.5$ and two at $\pm1$. The height of the maximum at the origin is $h_0=a/18$, and the other two are at $\pm0.79$; they are $h_{\pm}$, and are $3.16$ times shorter. Moreover, the derivative of the potential to be used in Eq. \eqref{var2} has the form
\be  
\frac{dV(x_n)}{dx_n}=\frac29\, a \, x_n\,( 32\,x^6_n -60\,x^4_n + 33\,x^2_n - 5 ).
\ee
This potential describes a much more involved situation, with the small particles having a more complex internal structure. In the same way as the fourth-order model, we can note the fixed point at the origin is stable for small values of $a$. For example, in the Fig. \ref{fig3}a, we observe orbits for $a=0.5$, which engenders the elliptical structure centered on the fixed point $(0,0)$, but we also see the emergence of new elliptical structures related to the other two maxima of the potential. The same appears in Fig. \ref{fig3}b, for $a=1.83$, but now the orbits are different.

\begin{figure}
	\includegraphics[width=8cm]{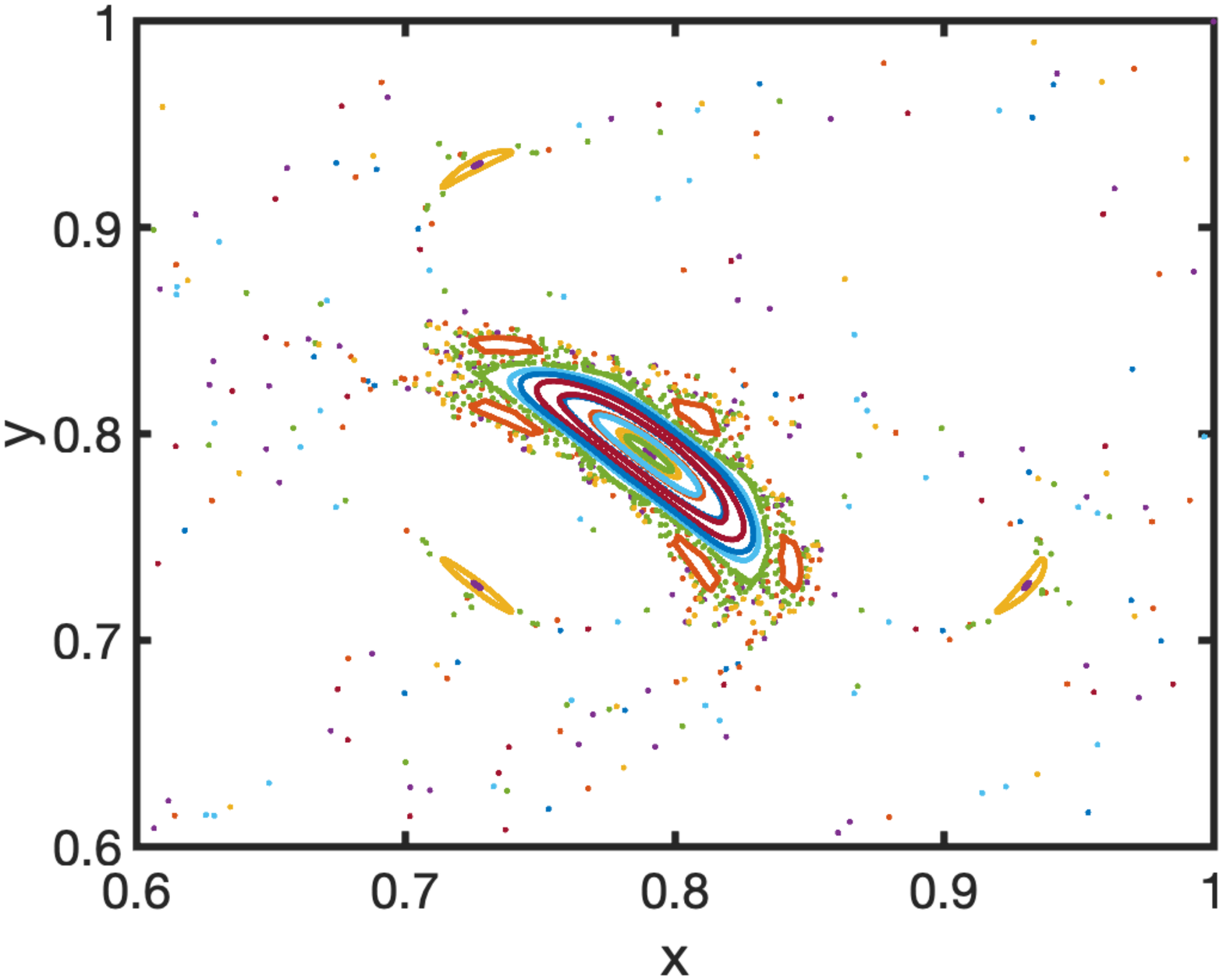} 
	\includegraphics[width=8cm]{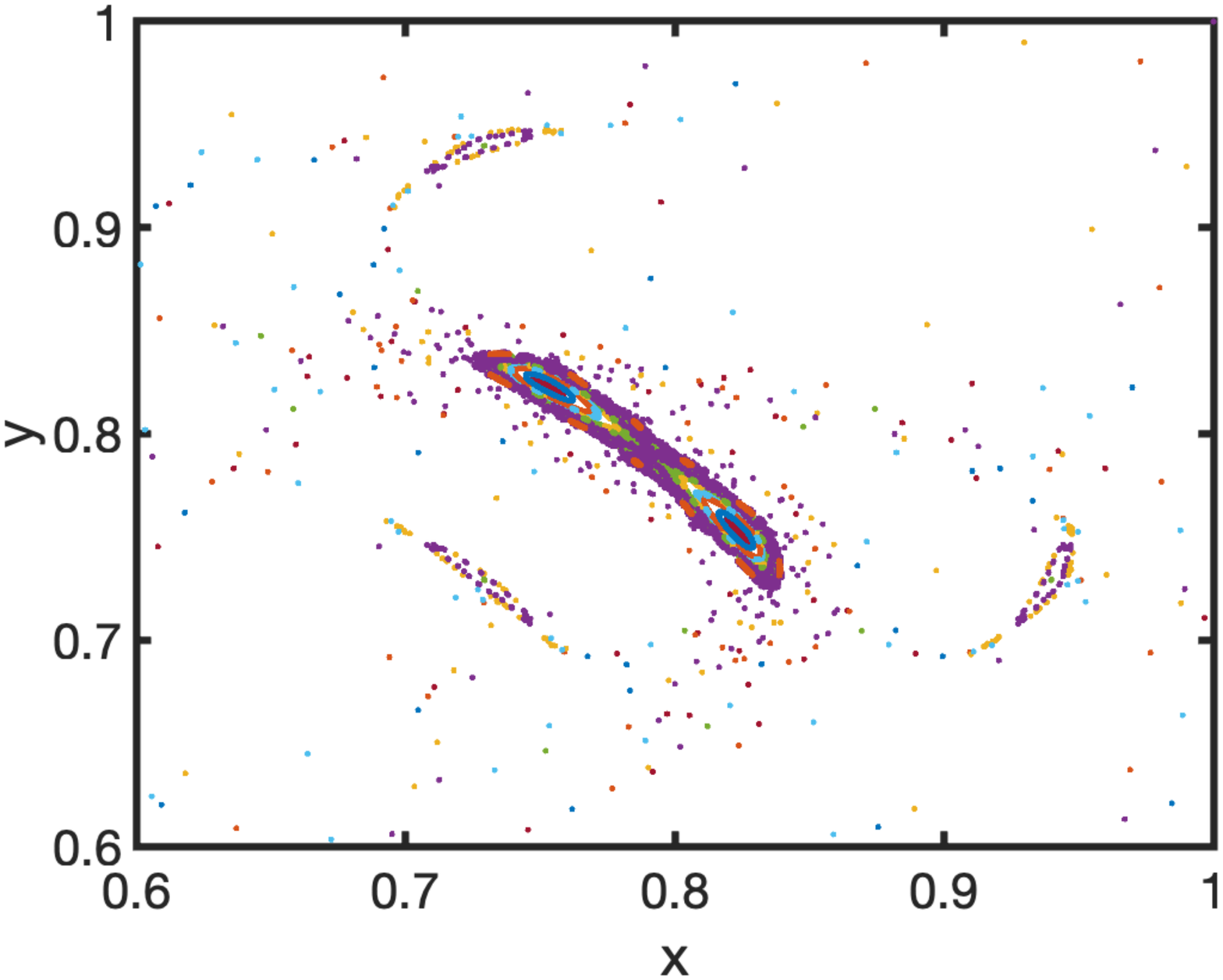} 
	\caption{(a) Orbits of the eighth-order model calculated for $a=3.1$ and (b) $a=3.3$.}
	\label{fig4}
\end{figure}

\begin{figure}
	\includegraphics[width=8cm]{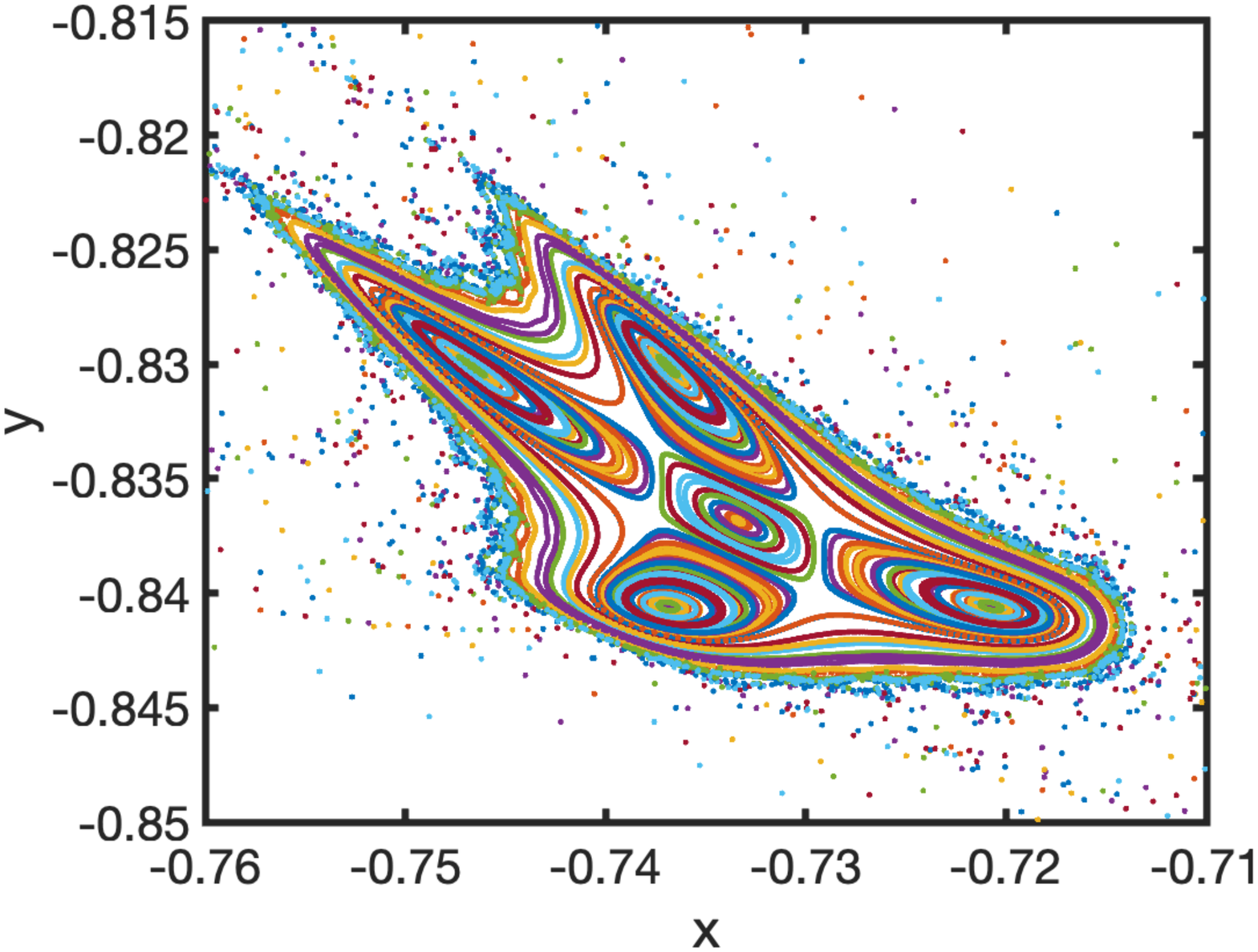} 
	\includegraphics[width=8cm]{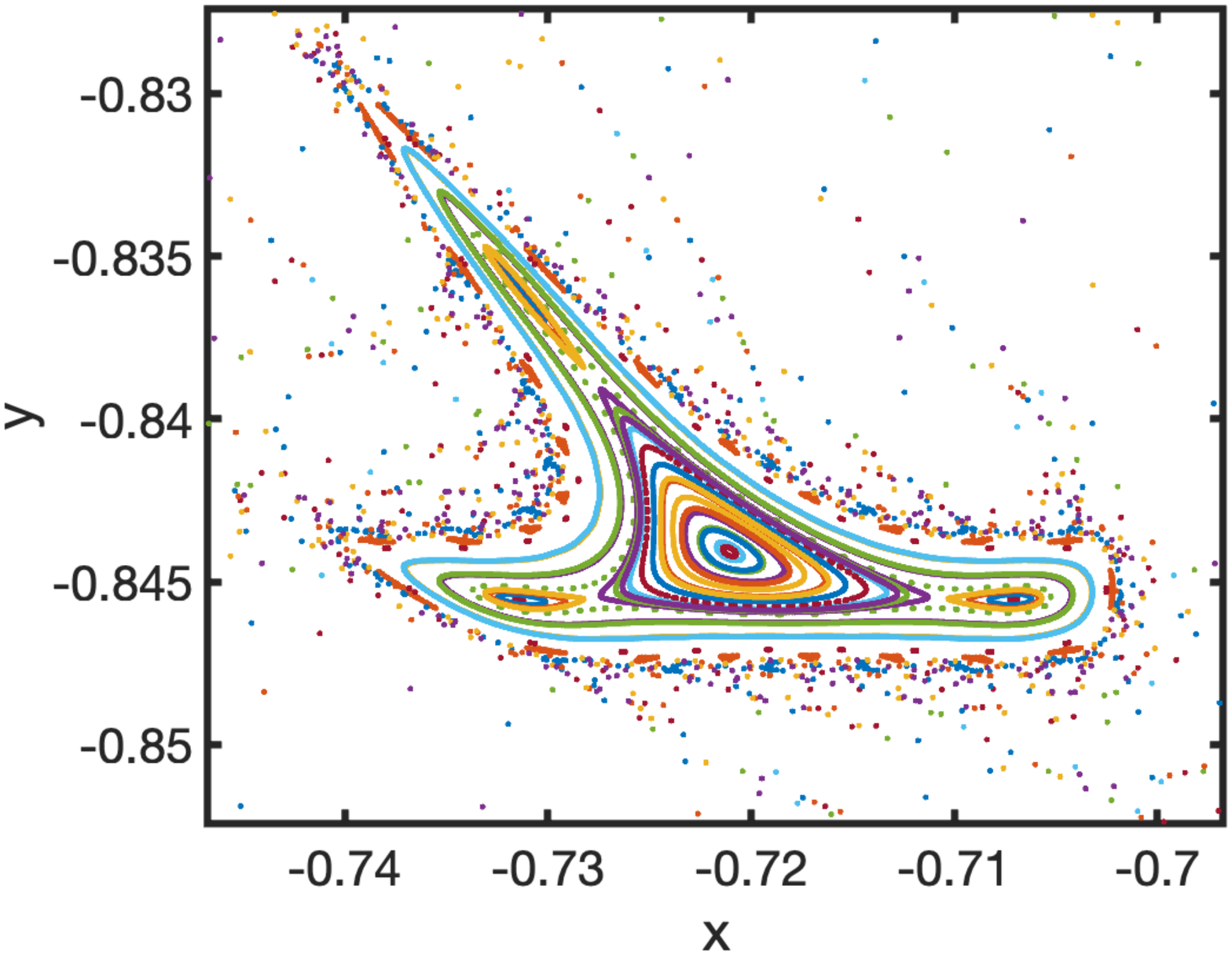} 
	\caption{(a) Orbits of the eighth-order model calculated for $a=3.422$ and for (b) $a=3.524$, showing the starfish and the three-armed star, respectively.}
	\label{fig4B}
\end{figure}

\begin{figure}
	\includegraphics[width=8cm]{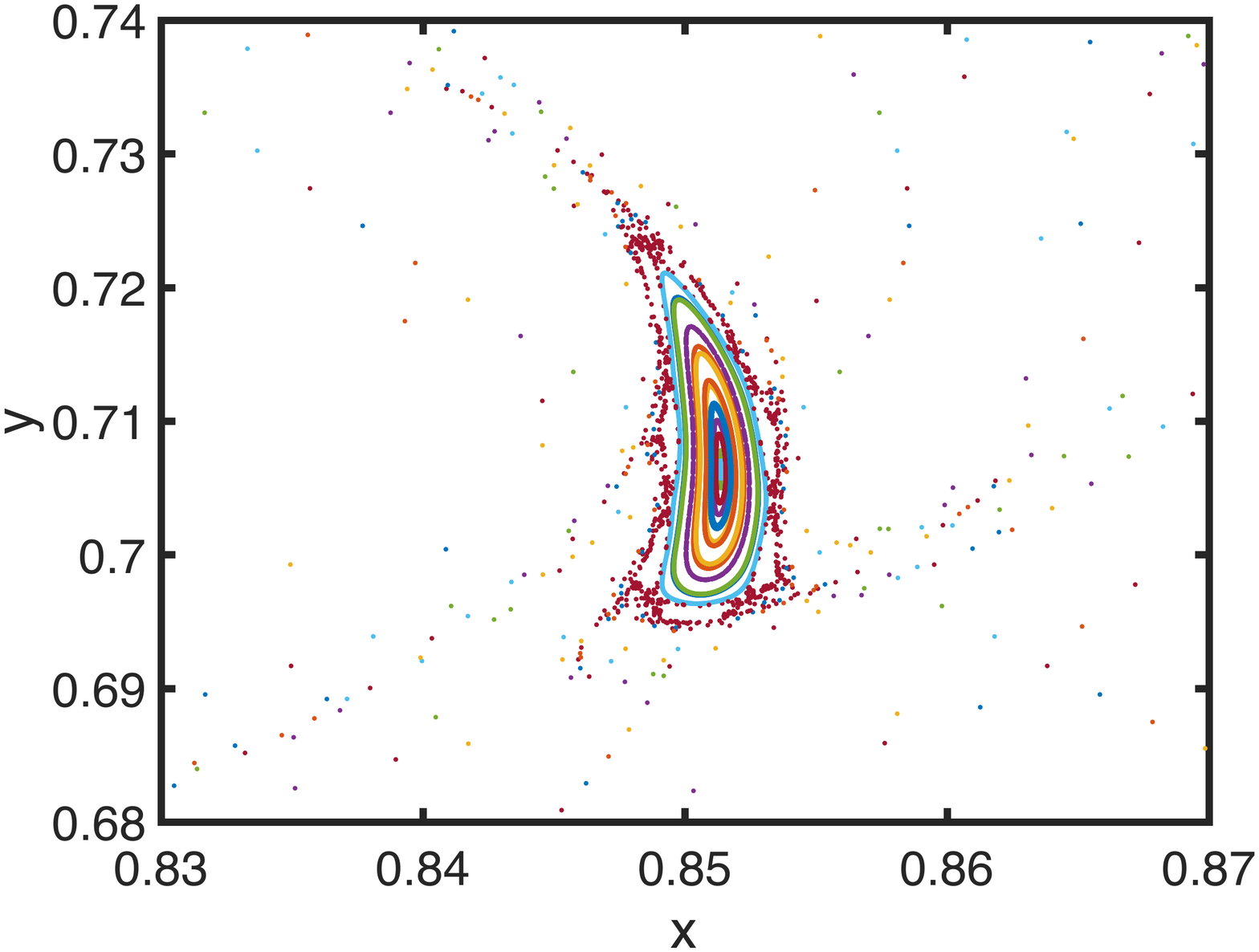} 
	\includegraphics[width=8cm]{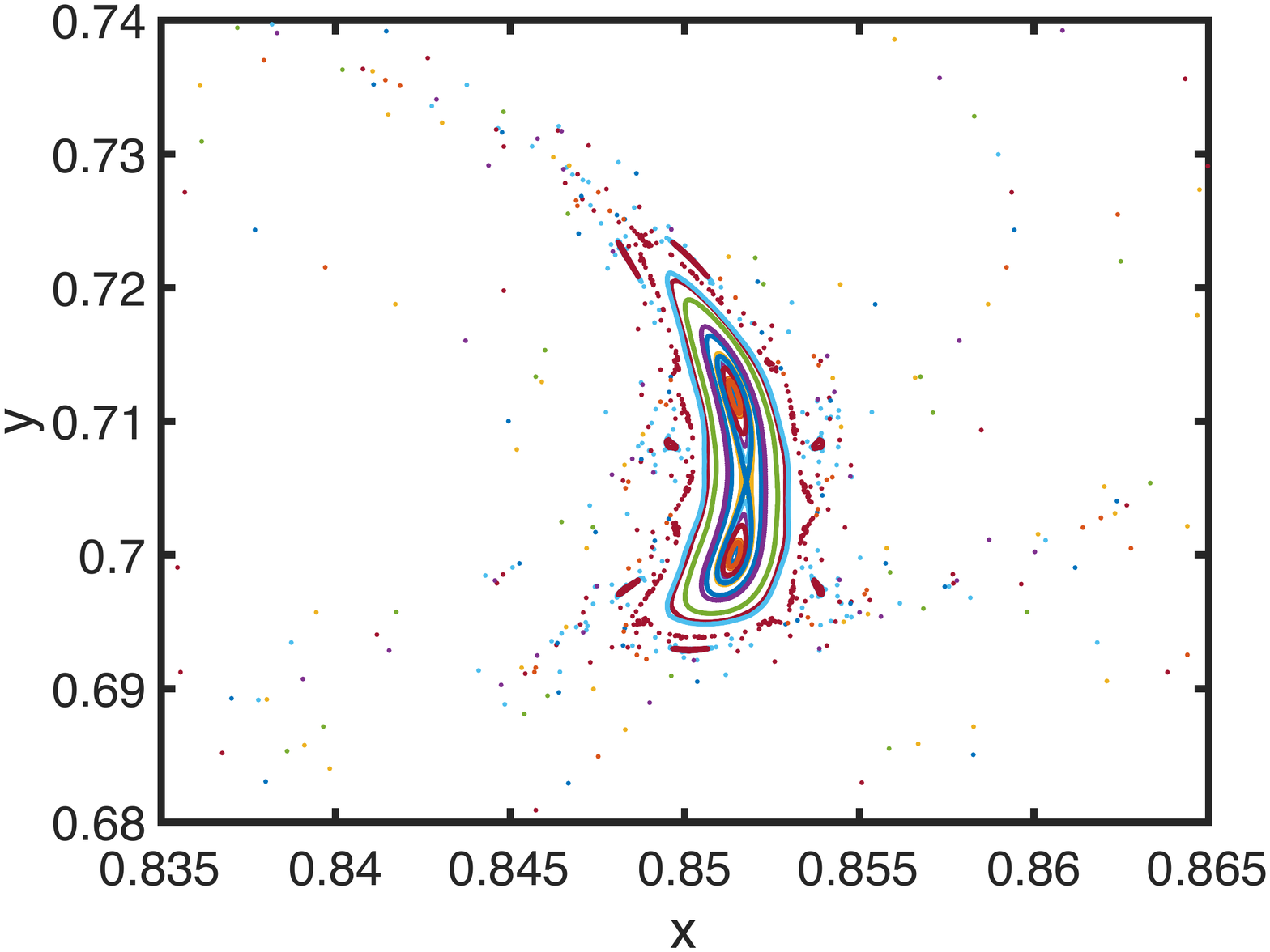} 
	\includegraphics[width=8cm]{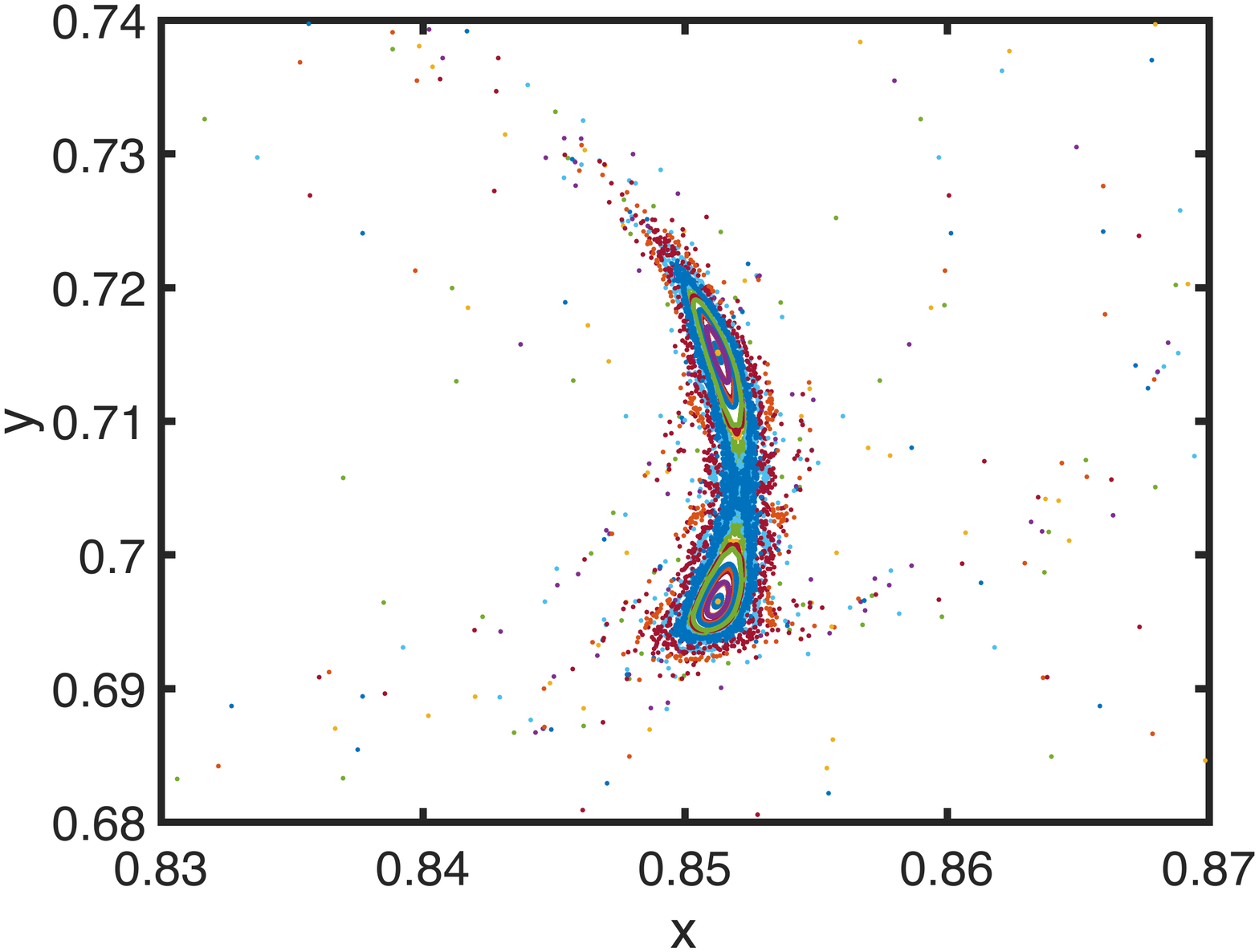} 
	\caption{ The banana behavior of the eighth-order model calculated for (a) $a=3.672$, (b) $a=3.682$ and (c)  $a=3.69$.}
	\label{fig4C}
\end{figure}

Another intriguing phenomenon is seen with the increase in $a$. We can see in Fig. \ref{fig3}b that the new orbits migrate away from the core of the fixed points. The orbits begin to rotate and change their elliptical form. This behavior is directly related to the first minimum point found in the Lyapunov exponent, which will be addressed later. It is worth noting that the orbits around $(0,0)$ and $\pm(0.79,0.79)$ behave similarly, distorting their initial elliptical shape. Following the Refs. \cite{bak1,bou1,gree1}, we can analyze the stability of the fixed points through the tangent space orbits, as done before for the fourth-order potential. In comparison to the previous model, the eighth-order potential is more involved. For this reason, let us now focus on the behavior of the model around the maximum at $(0.79,0.79)$. It bifurcates at $a=3.2$, so in Fig. \ref{fig4} we depict the orbits for $a=3.1$ and $a=3.3$, which illustrate the corresponding behavior. We increase the value of $a$, searching for other possibilities.
We have also found the starfish, the three armed star and the banana behavior. They are displayed in Fig. \ref{fig4B} and \ref{fig4C}, and show behavior similar to the case of the fourth-order potential, but this happens here in the interval $a\in [3.42,3.70]$ three times narrower than it appear in the previous model.

\begin{figure}
	\includegraphics[width=8cm]{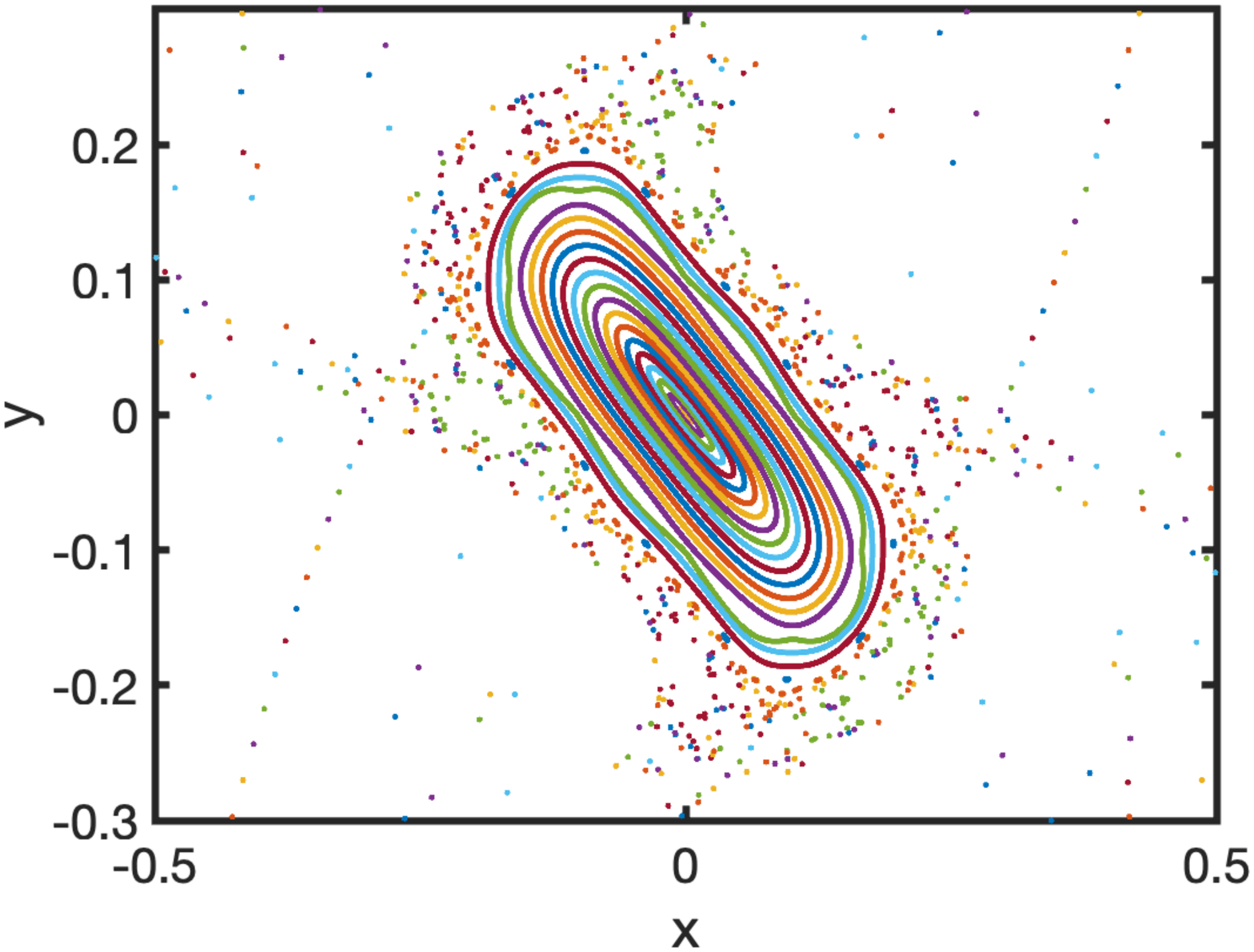} 
	\includegraphics[width=8cm]{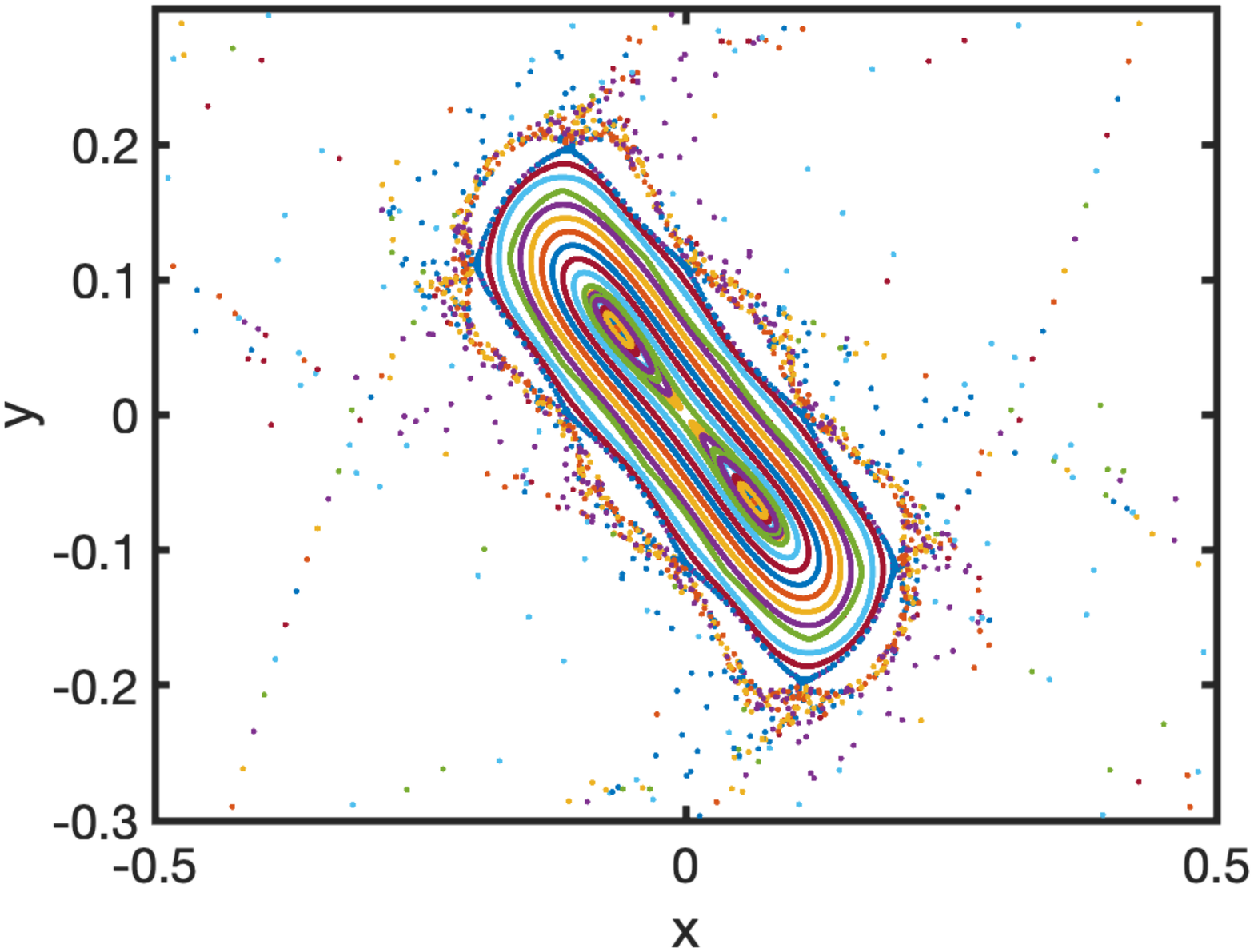}
	\caption{(a) Orbits of the eighth-order model calculated for $a=3.5$ and (b) $a=3.7$.}
	\label{fig4A}
\end{figure}

We now focus on the fixed point at the origin. We implemented numerical investigation and noted that it becomes unstable for $a>3.6$. In the Fig. \ref{fig4A}, we can see the orbits for $a=3.5$ and $a=3.7$, respectively, and they clearly show the corresponding bifurcation. When comparing Fig. \ref{fig4A} to Fig. \ref{fig3}, we can also see that the orbits have different inclination.

\begin{figure}
	\includegraphics[width=8cm]{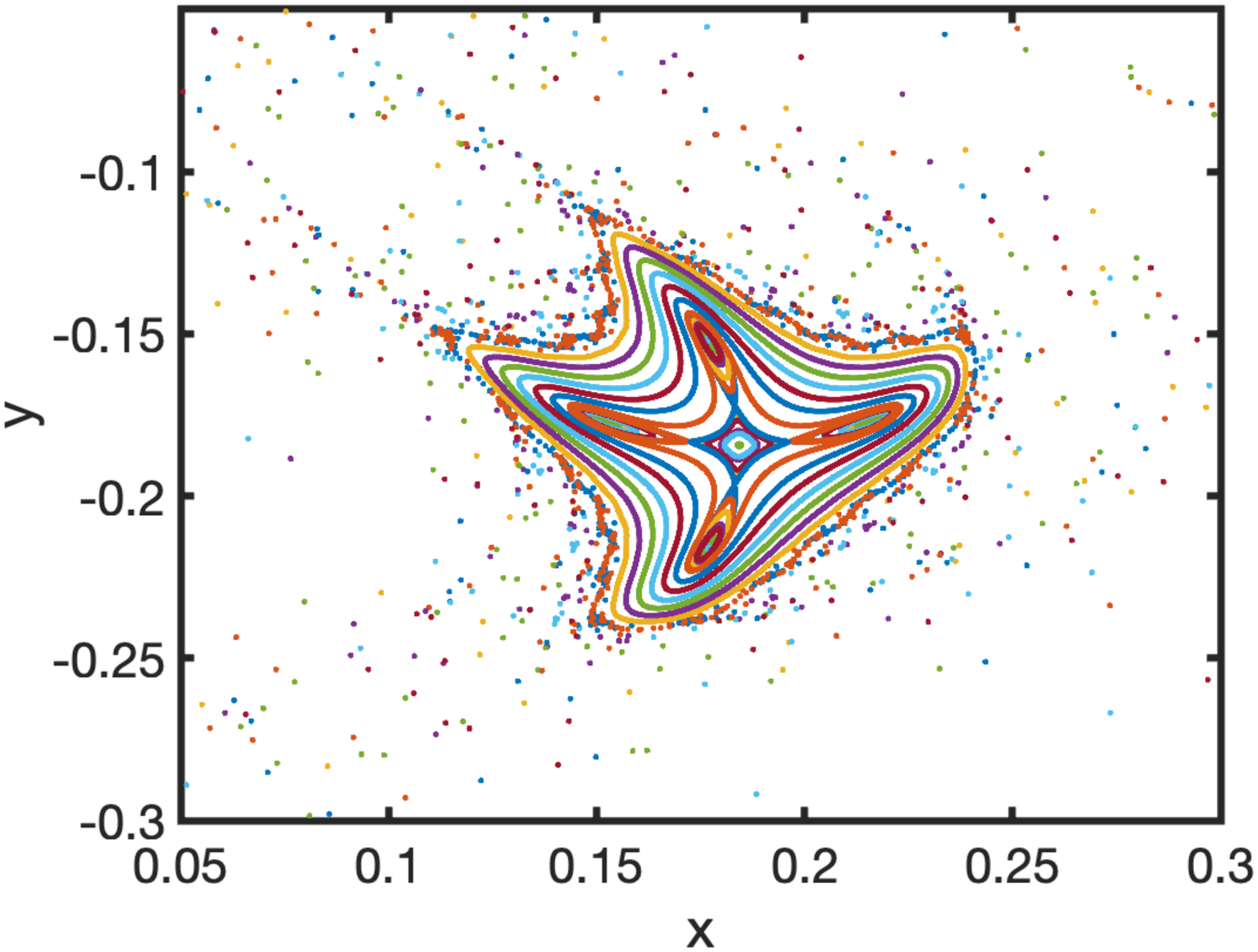}
    \includegraphics[width=8cm]{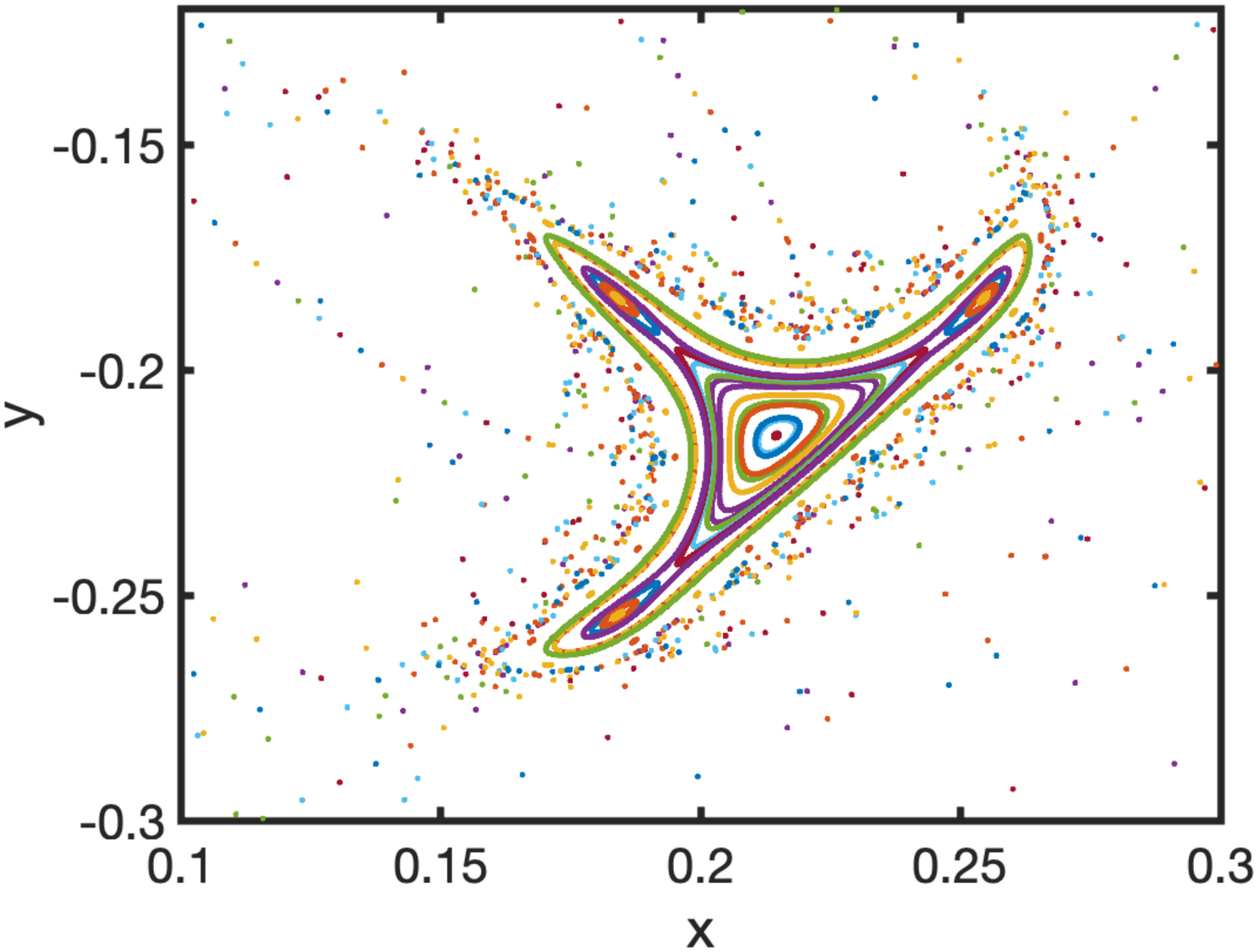} 
	\caption{(a) Orbits of the eighth-order model calculated for $a=4.56$ and (b) $a=4.993$, showing the starfish and the three-armed star, respectively.}
	\label{fig7}
\end{figure}

\begin{figure}
	\includegraphics[width=8cm]{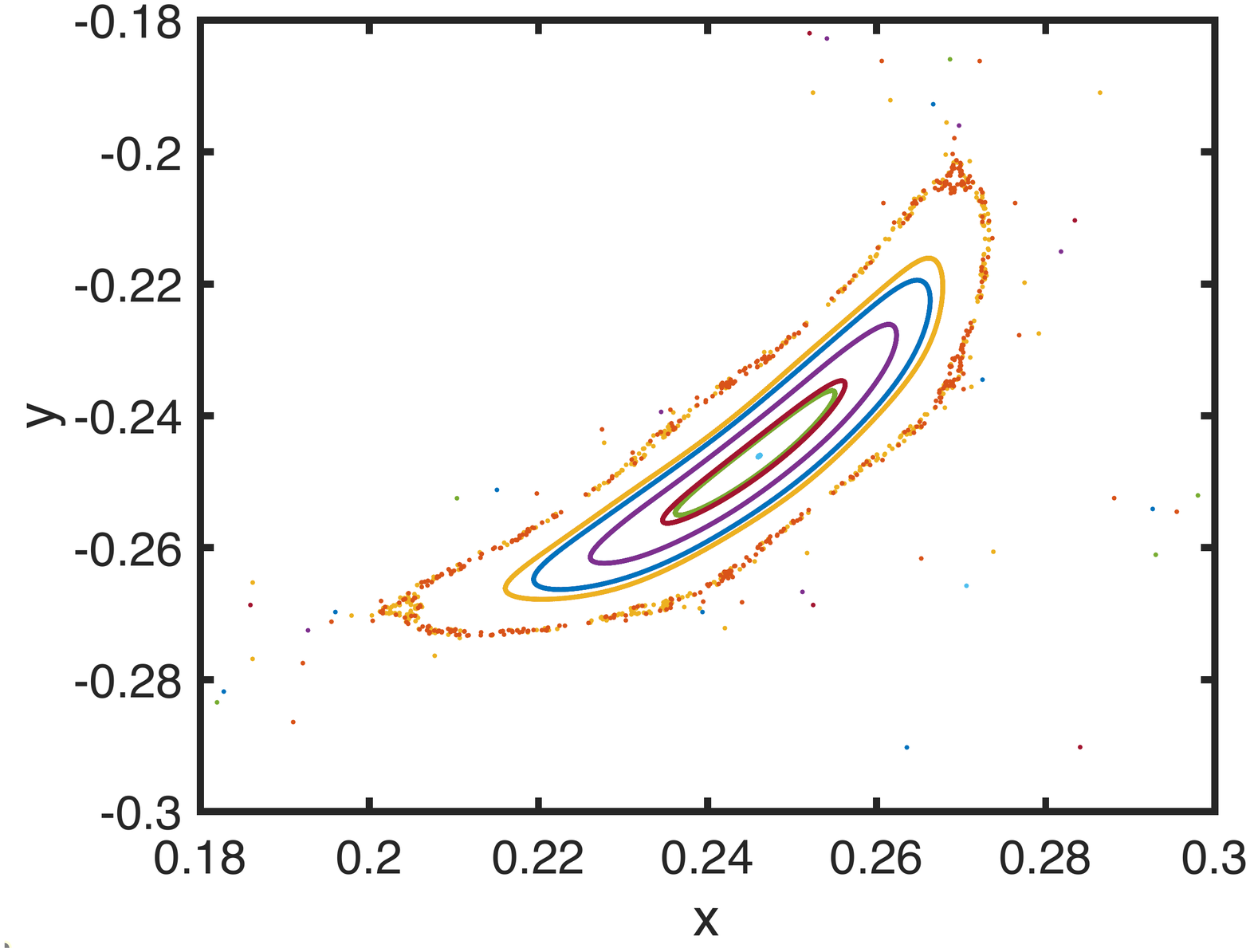} 
	\includegraphics[width=8cm]{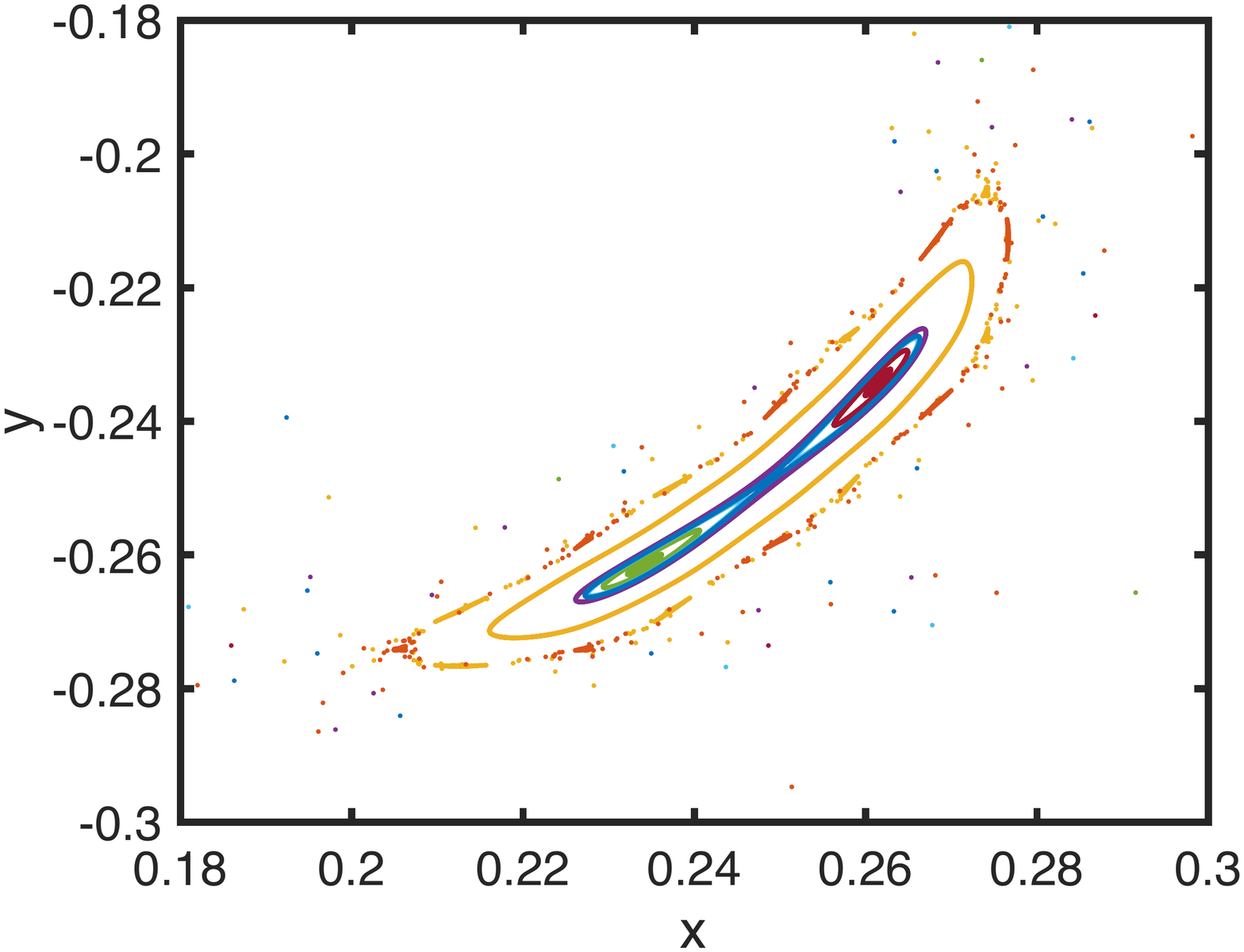}
	\includegraphics[width=8cm]{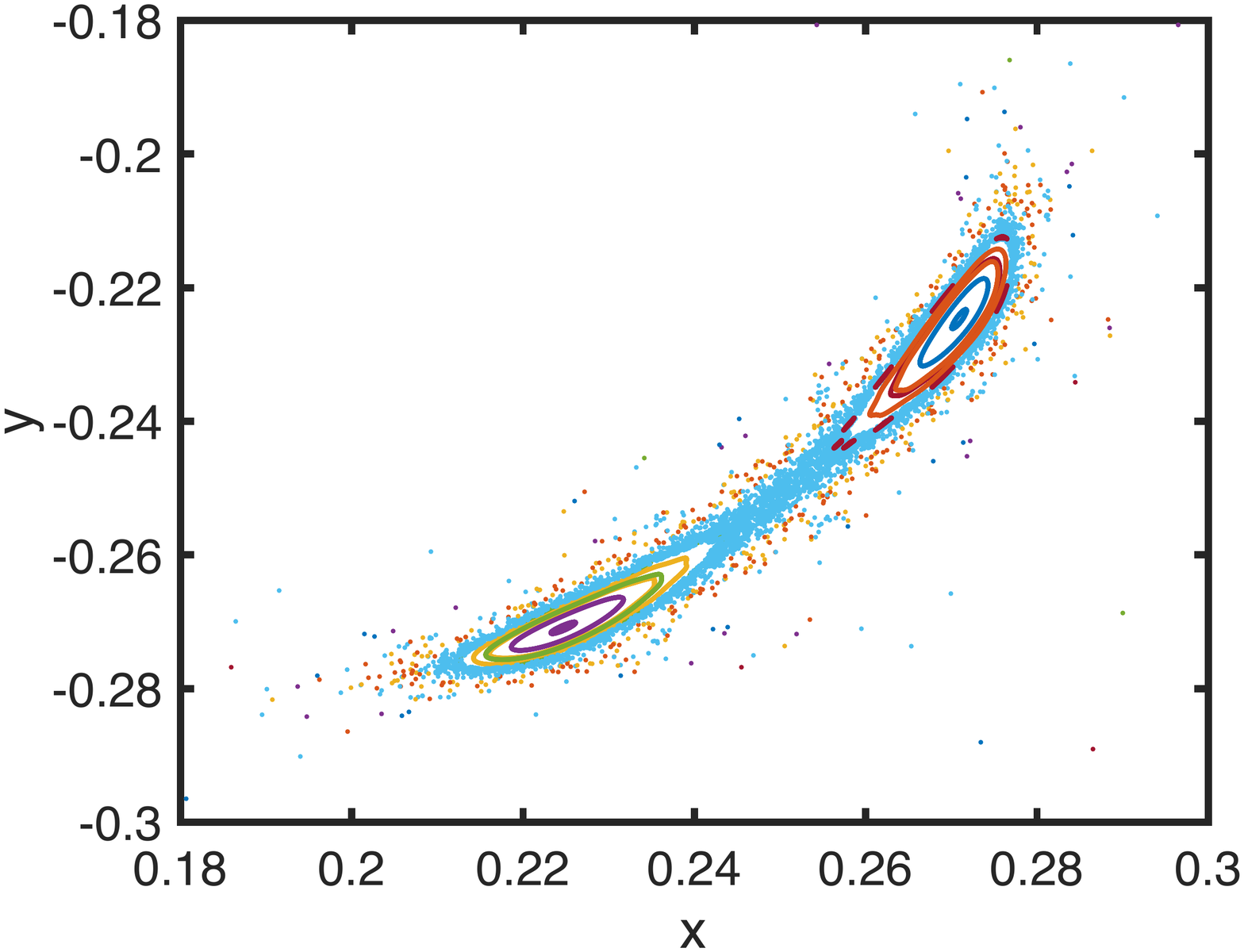}
	\caption{The banana behavior of the eighth-order model calculated for (a) $a=5.6$, (b) $a=5.66$ and (c) $a=5.7$.}
	\label{fig5}
\end{figure}

We increase the value of $a$ and in Fig. \ref{fig7}a we display the starfish behavior for $a=4.56$. This is similar to the fourth-order model, so we increase even more $a$ until we find in Fig. \ref{fig7}b the three-armed star, and then in Fig. \ref{fig5} we depict the banana behavior for several values of $a$. These results show that for $a$ in the interval from $a=4.5$ until $a=5.7$, the eighth-order model behaves as the previous model, with the fourth-order potential, showing the starfish, the three-armed star and the banana behavior as well.

\begin{figure}
	\includegraphics[width=8cm]{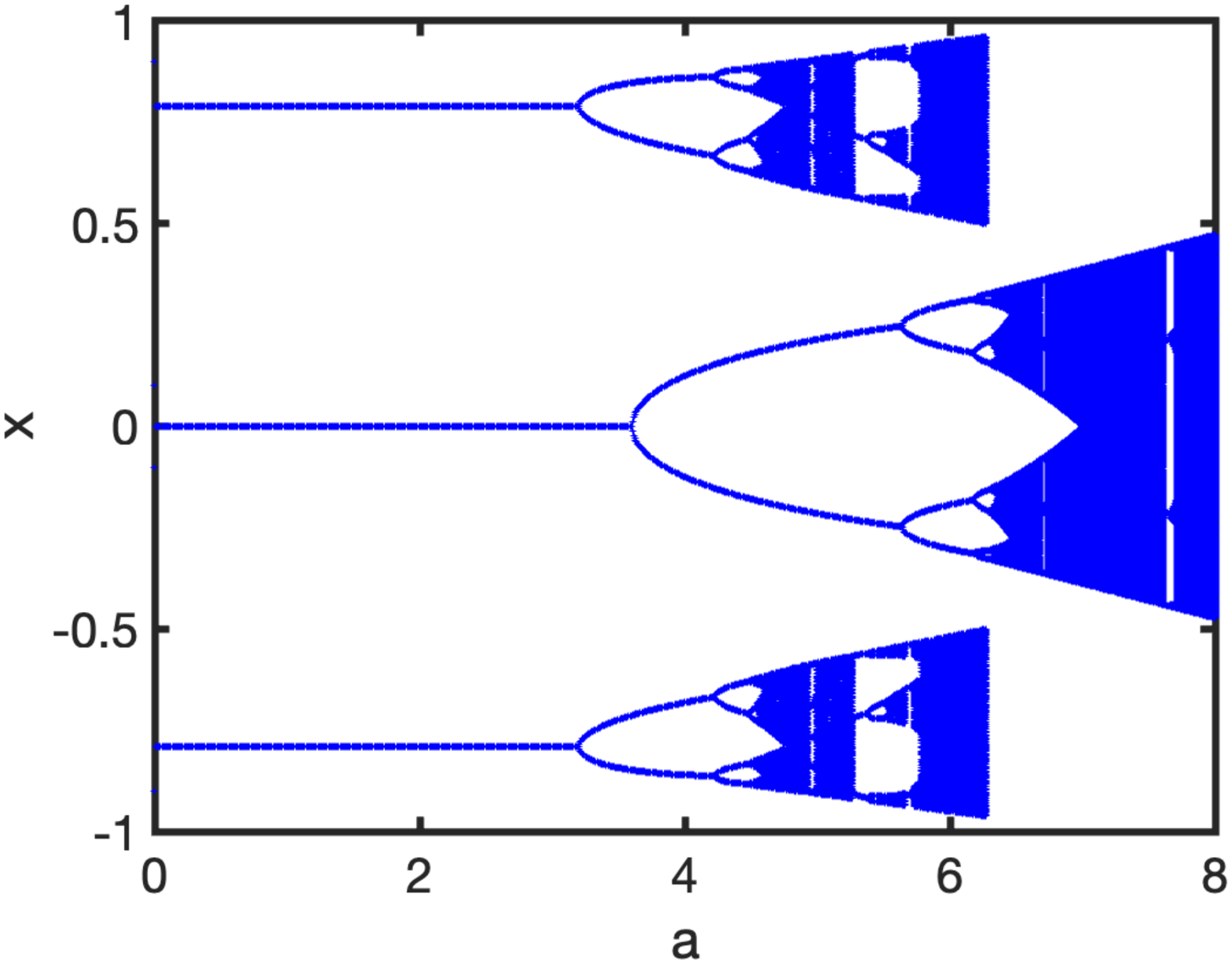} 	
	\includegraphics[width=8cm]{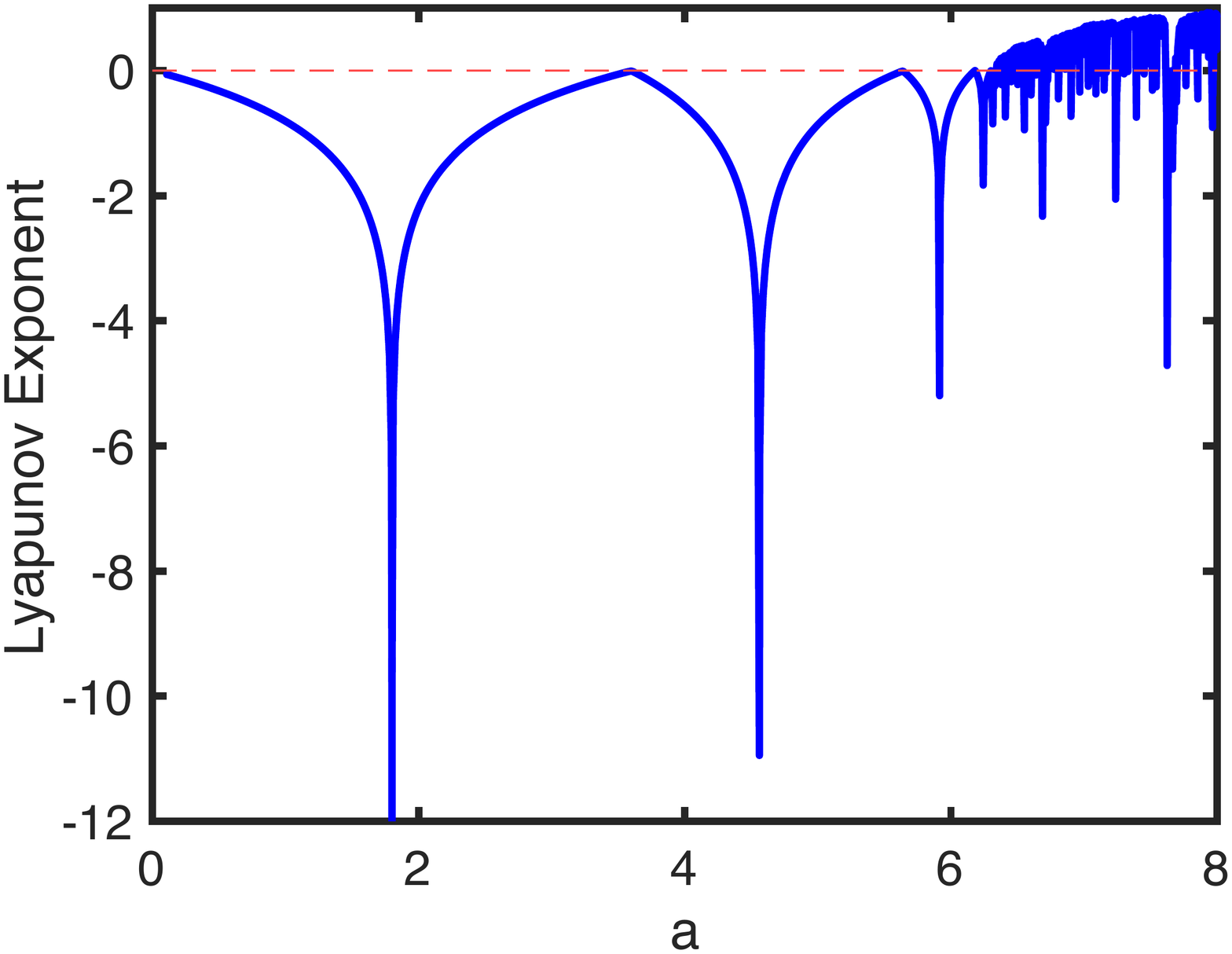} 
	\caption{(a) Bifurcation diagram of the eighth-order model and (b) corresponding behavior of Lyapunov exponent.}
	\label{fig6}
\end{figure}

We have added more investigation, and in Fig. \ref{fig6}a one depicts the bifurcation behavior of the model as the value of the parameter $a$ increases. The diagram shows the path of new orbit formation until the dynamic of the system becomes chaotic. As we can see, the bifurcation process continues for larger and larger values of $a$, and leads to chaos. In order to reinforce this behavior, in Fig. \ref{fig6}b, we calculate the Lyapunov exponent for the eighth-order model, in order to identify the chaotic behavior. We did this following the behavior associated to the maximum at the origin, in a way similar to the case described in Fig. \ref{lya}b for the fourth-order potential. For $a<a_\infty=6.333$, we identify the presence of the values $a=3.598$, $a=5.637$ and $a=6.181$ where the Lyapunov exponent approaches zero, also showing the first, second and third bifurcation, in direct connection with the corresponding bifurcation diagram. Furthermore, for large values of $a$, the Lyapunov exponent becomes positive, indicating that the system has reached the chaotic behavior. It is also worth noting that the exponent has minima at $a=1.8$ and $a=4.56$, for instance. The first minimum relates to the limit region in which elliptical orbits begin to rotate at $90^{\circ}$, as compared to smaller values of $a$. The second minimum refers to a region in which the 2-cycles become marginally unstable, but no bifurcation is observed. The usual starfish behavior can be found in this instability zone, as we can see in Fig. \ref{fig7} for $a=4.56$.

In order to help us compare the results on bifurcation and the Lyapunov exponent in the fourth- and eighth-order models, we have depicted both Figs. \ref{lya}a and \ref{fig6}a with the same number of points, and Figs. \ref{lya}b and \ref{fig6}b with curves with the same width. As one can see, the fourth-order model is governed by a double well potential, and the eighth-order model by a much more involved potential. The two models are different from each other, and develop distinct bifurcation behavior.

\section{Conclusion}\label{sec3}

In this work we have revisited the investigations described in Refs. \cite{bak1,bak2}, concerning an one dimensional chain of small particles, which interact via the fourth-order potential displayed in Eq. \eqref{p4}. As shown in \cite{bak2}, the non-linearity introduced by the particles may induce the system to exhibit bifurcation and chaos. Moreover, in the more recent investigation in \cite{magnetic}, one also sees the appearance of fractal structures in the one dimensional chain of small magnetic particles.

The interesting behavior has attracted our attention, and here we considered the study of the Lyapunov exponent, to see how it fits with the orbits and the associated bifurcation pattern. The new results presented in Fig. \ref{lya} for the fourth-order potential unveiled interesting relation between the bifurcation diagram and the Lyapunov exponent, motivating us to go further and study other possibilities. We have, in particular, found in between the starfish and the banana behavior, the new three-armed star that appears in Fig. \ref{satr2}b, which was not present in previous investigations. We have also considered the fourth-order model with the inclusion of a new parameter, $b$, to control the fourth-order interaction, and another model, with another potential. This is the eighth-order potential displayed in Eq. \eqref{p8}, which has a more complex structure and may be used to describe one dimensional chain of small particles engendering other kind of non-linearity. The results reported in this work, in particular the ones for bifurcation and the Lyapunov exponent displayed in Fig. \ref{fig6}  show that modifications in the non-linearity related to the small particles in the chain changes the behavior of the system. The two systems are different, and we can see this from the two Figs. \ref{lya} and \ref{fig6}, so they suggest that we further investigate other possibilities. 

An interesting specific possibility of investigation could follow Ref. \cite{magnetic}, to see how the fractal structure appears in this new situation, with the model being described by the eighth-order potential considered in the present work. Another possibility could be the case where the distinct interactions are controlled by distinct real parameters, which weight their importance in the model, to see how the non-linearities contribute to induce or generate the chaotic behavior in the chain. We can also suppose that in the one dimensional chain of small particles, the particles interact with the nearest-neighbour and also with the next-nearest-neighbour. The presence of the next-to-nearest interaction will certainly change the kinematics of the model and may perhaps bring new information on the behavior of the system. We can also suppose that the small particles are composed, described by two (or more) distinct degrees of freedom, to see how they can contribute to the chaotic behavior. This is motivated by investigations developed before, on the presence of localized structures in relativistic models described by coupled real scalar fields; see, e.g., Ref. \cite{Baz,2fields1,Juan,2fields2} and references therein for more information on this issue. In particular, in \cite{2fields2} the study focuses on bifurcation and pattern changing. Evidently, the presence of composed particles increases the possibility of applications of practical importance.

There are several other problems of current interest, in particular, the ones concerning changing the one dimensional chain of small particles and consider a two dimensional lattice where several species evolve under cyclic and non hierarchical rules similar to the rules of the rock, paper and scissors game. Studies concerning the chaotic behavior of this kink of system were implemented before in Refs. \cite{srep,EPL}, for instance, based on the Hamming distance concept, but now we want to investigate the Lyapunov exponent. There are several investigations dealing with this issue  in predator-pray systems in a two dimensional lattice, and we think the works \cite{AA,BB} will certainly help us implement other possibilities.

We can also suppose that the one dimensional chain of small particles is composed of atomic elements. This may bring quantum effects and change radically the models investigated in this work. However, it is of current interest and here we recall, for instance, the case of quantum particles on a chain of harmonically coupled quartic double-well potential \cite{prb} and also, the systems composed of two distinct family of atomic bosonic and fermionic particles, the one dimensional optical lattices of $^7$Li -- $^6$Li and $^{23}$Na -- $^6$Li investigated before in Refs. \cite{A,B}. We are now considering some of the above possibilities, and hope to report on them in the near future.


\section{Acknowledgements}


This work is supported by funds provided by Conselho Nacional de Desenvolvimento Cient\'ifico e Tecnol\'ogico, CNPq, Grant no. 303469/2019-6, and by Paraiba State Research Foundation, FAPESQ-PB, Grant no. 0015/2019. F.C.S. thanks FAPEMA, Funda\c c\~ao de Amparo \`a Pesquisa e ao Desenvolvimento Cient\'ifico e Tecnol\'ogico do Maranh\~ao, Universal Grant 00920/19.

\begin{center}{*****}\end{center}

The authors declare that there are no known conflicts of interest associated with this publication and there has been no financial support for this work that could have influenced its outcome.\\

\begin{center}{*****}\end{center}

DB suggested the investigation. DB, KZN and FCS discussed the analytical calculations and the organization of the work. KZN and FCS implemented the numerical calculations. DB and FCS selected the main results. DB wrote the paper and KZN and FCS agreed with the final version. 


\end{document}